\documentclass[singlecolumn,epjc3]{svjour3}
\smartqed  
\RequirePackage{graphicx}
\journalname{Eur. Phys. J. C}

\begin{document}

\title{Charged Perfect Fluid Sphere in Higher Dimensional Spacetime}

\author{Piyali Bhar\thanksref{e1,addr1}
\and Tuhina Manna\thanksref{e2,addr2} \and Farook
Rahaman\thanksref{e3,addr3} \and Saibal Ray\thanksref{e4,addr4}
\and G.S. Khadekar\thanksref{e5,addr5} .}

\thankstext{e1}{e-mail: piyalibhar90@gmail.com}
\thankstext{e2}{e-mail: tuhinamanna03@gmail.com}
\thankstext{e3}{e-mail: rahaman@iucaa.ernet.in}
\thankstext{e4}{e-mail: saibal@iucaa.ernet.in}
\thankstext{e5}{e-mail: gkhadekar@rediffmail.com}

\institute{Gopinathpur High School (H.S), Haripal, Hooghly 712403,
West Bengal, India\label{addr1} \and Department of Mathematics,
Jadavpur University, Kolkata 700032, West Bengal,
India\label{addr2} \and Department of Commerce (Evening), St.
Xaviers Colege, 30 Mother Teresa Sarani, Kolkata 700016, West
Bengal, India\label{addr3} \and Department of Physics, Government
College of Engineering \& Ceramic Technology, 73 A.C.B. Lane,
Kolkata 700010, West Bengal, India\label{addr4} \and Department of
Mathematics, R.T.M. Nagpur University, Mahatma Jyotiba Phule
Educational Campus, Amravati Road, Nagpur 440033, Maharastra,
India\label{addr5}}

\date{Received: date / Accepted: date}

\maketitle

\begin{abstract}
Present paper provides a new model for perfect fluid sphere filled with charge
in higher dimensional spacetime admitting conformal symmetry.
We consider a linear equation of state with coefficients fixed by the
matching conditions at the boundary of the source corresponding to
the exterior Reissner-Nordstr\"{o}m higher dimensional spacetime.
Several physical features for different dimensions,
starting from four up to eleven, are briefly discussed. It is shown that 
all the features as obtained from the present model are physically desirable
and valid as far as the observed data set for the compact star $SAX~J1808.4-3658~(SS2)$ is concerned. 
\end{abstract}


\section{Introduction}
With the recent advancement in  superstring theory in which the 
spacetime is considered to be of dimensions higher than four, 
the studies in higher dimensional spacetime has attained new importance. 
Throughout the last decade a number of articles have been published 
in this subject both in localized and cosmological domains. It is a 
common trend to believe that the $4$-dimensional present 
spacetime structure is the self-compactified form of manifold with 
multidimensions. Therefore, it is argued that theories of unification 
tend to require extra spatial dimensions to be consistent with the 
physically acceptable models~\cite{Schwarz1985,Weinberg1986,Duff1995,Polchinski1998,Hellerman2007,Aharony2007}. 
It has been shown that some features of higher dimensional black holes 
differ significantly from $4$-dimensional black holes as 
higher dimensions allow for a much richer landscape of black hole 
solutions that do not have 4-dimensional 
counterparts~\cite{Emparan2008}. Some recent higher dimensional works 
admitting one parameter Group of Conformal motion can be seen 
in the Refs.~\cite{Pradhan2007,Khadekar2014}. 

The study of charged fluid sphere has attained considerable interest among 
researchers in last few decades. It is observed that a fluid sphere of 
uniform density with a net surface charge becomes more stable than without
charge~\cite{stettner}. According to Krasinski~\cite{Krasinski} 
in the presence of charge, the gravitational collapse of a spherically 
symmetric distribution of matter to a point singularity may be avoided. 
Sharma et al.~\cite{sharma01} argue that in this situation the repulsive 
Colombian force counterbalances the gravitational attraction in addition 
to the pressure gradient. To study the cosmic censorship hypothesis and the
formation of naked singularities Einstein-Maxwell solutions are
also important~\cite{joshi}. The presence of charge affects the
values for redshift, luminosity and maximum mass for stars. For a
charged fluid spheres the gravitational field in the exterior
region is described by Reissner-Nordstr\"{o}m spacetime. Charged
perfect fluid sphere satisfying a linear equation of state was
discussed by Ivanov~\cite{ivanov}. In this paper the author
reduced the system to a linear differential equation for one
metric component. Regular models with quadratic equation of state
was discussed by Maharaj and Takisa~\cite{maharaj13}.

The obtained solutions of the Einstein-Maxwell system of equations
are exact and physically reasonable. A physical analysis of the
matter and electromagnetic variables indicates that the model is
well behaved and regular. In particular there is no singularity in
the proper charge density at the stellar center. A Charged
anisotropic matter with linear equation of state has discussed by
Thirukkanesh and Maharaj~\cite{maharaj08}. In connection with this
we want to mention a recent work of Varela et al.~\cite{verela}. 
In this paper the author considered a self-gravitating, charged 
and isotropic fluid sphere.

To solve Einstein-Maxwell field equation they have assumed both
linear and nonlinear equation of state and discussed their result
analytically. Rahaman et al.~\cite{frahaman10} have obtained
a singularity free solutions for anisotropic charged fluid sphere
with Chaplygin equation of state. The authors used Krori-Barua 
ansatz~\cite{kb} to solve the system.

The well known inheritance symmetry is the
symmetry under conformal killing vectors (CKV) i.e.
\begin{equation}
L_\xi g_{ik}=\psi g_{ik},
\end{equation}
where $L$ is the Lie derivative of the metric tensor which
describes the interior gravitational field of a compact star with
respect to the vector field $\xi$ and $\psi$ is the conformal
factor. In a deeper sense this inheritance symmetry provides the 
natural relationship between geometry and matter through the 
Einstein field equations. It is supposed that the vector $\xi$ generates the
conformal symmetry and the metric $g$ is conformally mapped onto
itself along $\xi$. Harko et al.~\cite{Harko07,Harko08} have shown 
that neither $\xi$ nor $\psi$ need to be static
even through one consider a static metric.

There are many earlier works on conformal motion in literature.
The existence of one parameter group of conformal motion in
Einstein-Maxwell spacetime have been studied
in~\cite{krori86a,krori86b,maartens}. Anisotropic sphere admitting
one-parameter group of conformal motion has been discussed by
Herrera and Le\'{o}n~\cite{leon85a}. A class of solutions for
anisotropic stars admitting conformal motion has been studied in
\cite{farook10}. Charged gravastar admitting conformal motion has
been studied by Usmani {\emph et al.}~\cite{usmani11}. Bhar~\cite{bhar1} 
has generalized this result in higher dimensional spacetime. 
Relativistic stars admitting conformal motion has
been analyzed by Rahaman et al.~\cite{rahaman10}. Isotropic and
anisotropic charged spheres admitting a one parameter group of
conformal motions was analyzed in \cite{leon85b}. Anisotropic
spheres admitting a one parameter group of conformal motions has
been discussed by Herrera \& Le\'{o}n \cite{hleon85}. Charged
fluid sphere with linear equation of state admitting conformal
motion has been studied in Ref.~\cite{aloma10}. The authors have
also discussed about the dynamical stability analysis of the
system. Ray et al. have given an electromagnetic mass model
admitting conformal killing vector~\cite{ray04,ray07}. By assuming
the existence of a one parameter group of conformal motion Mak \&
Harko~\cite{mak04} have described an charged strange quark star
model. The above author have also discussed conformally symmetric
vacuum solutions of the gravitational field equations in the
brane-world models~\cite{hm05}. Bhar~\cite{Bhar2015a} has
described one parameter group of conformal motion in the presence of
quintessence field where the Vaidya-Titekar~\cite{vaidya} ansatz was
used to develop the model.

The obtained results are analyzed physically as well as with the
help of graphical representation. In a very recent work Bhar et
al.~\cite{Bhar2015b} provide a new class of interior solutions for
anisotropic stars admitting conformal motion in higher dimensional
noncommutative spacetime. The Einstein field equations are solved
by choosing a particular density distribution function of
Lorentzian type as provided by Nazari and Mehdipour
\cite{meh1,meh2} under a noncommutative geometry.

Inspired by these early works in the present paper we have used the
Einstein-Maxwell spacetime geometry to describe a self-gravitating charged
anisotropic fluid sphere satisfying a linear equation of state
admitting conformal motion in higher dimensions. Once we specify 
the equation of state (EOS) we have integrated the Tolman-Oppenheimer-Volkoff 
(TOV) equations to derive the gross features of the stellar
configuration. We propose to apply this model to describe charged
strange quark stars. The paper has been divided into the following
parts : In Sect. 2 we have obtained the Einstein-Maxwell field
equations for static spherically symmetric distribution of charged
matter. In Sect. 3 the conformal killing equations are solved 
and used the inheritance symmetry which is the symmetry under
conformal killing vectors (CKV). The exterior spacetime using
RN metric and investigation of the matching condition are also  
done here along with the matching of the exterior higher dimensional 
spacetime and interior spacetime at the boundary. In Sect. 4 various 
physical properties are analyzed such as (i) stability condition via 
the TOV equations are integrated to obtain the gravitational ($F_g$) 
and hydrostatic ($F_h$) forces, (ii) Energy conditions, namely, Null
energy condition (NEC), Weak energy condition (WEC) and Strong
energy conditions are discussed and the corresponding graphs for
different dimensions plotted against $r$ and (iii) the compactness
factor and redshift are investigated. Finally some concluding
remarks are passed in Sect. 5.

\section{The interior spacetime and Einstein-Maxwell Field equations}
To describe the static spherically symmetry spacetime in higher dimension we
consider the line element in the standard form as
\begin{equation}
ds^{2}=-e^{\nu(r)}dt^{2}+e^{\lambda(r)}dr^{2}+r^{2}d\Omega_n^{2},
\end{equation}
where
\begin{equation}
d\Omega_n^{2}=d\theta_1^{2}+\sin^{2}\theta_1d\theta_2^{2}+
\sin^{2}\theta_1\sin^{2}\theta_2d\theta_3^{2}+...+\Pi_{i=1}^{n-1}\sin^{2}\theta_id\theta_n^{2}
\end{equation}
and ${\lambda}$ and ${\nu}$ are functions of radial coordinate
$r$. Here dimension of the spacetime is assumed as $D=n+2$ so that
for $n=2$ it reduces to ordinary $4$-dimensional spacetime
geometry.

Now, the Einstein-Maxwell field equations in their fundamental forms
are given by
\begin{equation}
R_{ij} - \frac{1}{2}{g_{ij}} R = -\kappa ({T_{ij}}^{matter} + {T_{ij}}^{charge}) ,
\end{equation}
where the energy-momentum tensor of perfect fluid distribution is
\begin{equation}
{T_{ij}}^{matter} = (\rho + p) u_{i}u_{j} - p g_{ij},
\end{equation}
and the energy-momentum of the electromagnetic field is
\begin{equation}
{T_{ij}}^{charge} = \frac{1}{4\pi} [- g^{kl} F_{ik}F_{jl} + \frac{1}{4\pi} g_{ij}
F_{kl} F^{kl}].
\end{equation}
The electromagnetic field equations are given by
\begin{equation}
{[{(- g)}^{1/2} F^{ij}],}_{j} = 4\pi J^{i}{(- g)}^{1/2},
\end{equation}
and
\begin{equation}
F_{[ij,k]} = 0,
\end{equation}
where the electromagnetic field tensor $F_{ij}$ is related to the
electromagnetic potentials as $ F_{ij} = A_{i,j} - A_{j,i} $ which is
equivalent to the equation (6), viz., $ F_{[i,j,k]} = 0 $. Also,
$u^{i}$ is the 4-velocity of a fluid element, $J^{i}$ is the 4-current and $ \kappa = - 8 \pi $
(in relativistic unit $G = c = 1$). Here and in what follows a comma denotes
the partial differentiation with respect to the coordinate indices involving the index.

The Einstein-Maxwell field equations in higher dimension can be written as
\begin{equation}
e^{-\lambda}\left(\frac{n\lambda'}{2r}-\frac{n(n-1)}{2r^{2}}\right)+\frac{n(n-1)}{2r^{2}}=8\pi
\rho+E^{2},
\end{equation}

\begin{equation}
e^{-\lambda}\left(\frac{n(n-1)}{2r^{2}}+\frac{n\nu'}{2r}\right)-\frac{n(n-1)}{2r^{2}}=8\pi
p-E^{2},
\end{equation}

\begin{eqnarray}
\frac{1}{2}e^{-\lambda}\left[\frac{1}{2}(\nu')^{2}+\nu''-\frac{1}{2}\lambda'\nu'+\frac{(n-1)}{r}(\nu'-\lambda')
+\frac{(n-1)(n-2)}{r^{2}}\right] \nonumber \\-\frac{(n-1)(n-2)}{2r^{2}}=8\pi p
+E^{2}.
\end{eqnarray}

\begin{equation}
(r^{n}E)'=\frac{2\pi^{\frac{n+1}{2}}}{\Gamma\left({\frac{n+1}{2}}\right)}r^{n}\sigma(r)
e^{\frac{\lambda}{2}},
\end{equation}
where $\sigma(r)$ is the charge density on the $n$-sphere with
$n=D-2$,~$D$ being dimension of the spacetime.

The above equation equivalently gives
\begin{equation}
E(r)=\frac{1}{r^{n}}\int
\frac{2\pi^{\frac{n+1}{2}}}{\Gamma\left({\frac{n+1}{2}}\right)}r^{n}\sigma
e^{\frac{\lambda}{2}}dr,
\end{equation}
where $\rho,~p,~E$ are respectively the matter density, isotropic
pressure and electric field of the charged fluid sphere. Here
`prime' denotes the differentiation with respect to the radial
coordinate $r$.

\section{The solution under conformal Killing vector}
The conformal Killing equation (1) becomes
\begin{equation}
L_\xi g_{ik}=\xi_{i;k}+\xi_{k;i}=\psi g_{ik}.
\end{equation}

Now the conformal Killing equation for the above line element
$(2)$ gives the following equations
\begin{equation}
\xi^{1}\nu'=\psi,
\end{equation}

\begin{equation}
\xi^{n+2}=C_1,
\end{equation}

\begin{equation}
\xi^{1}=\frac{\psi r}{2},
\end{equation}

\begin{equation}
\xi^{1}\lambda'+2\xi^{1},_1=\psi,
\end{equation}
where $C_1$ is a constant.

The above equations consequently gives
\begin{equation}
e^{\nu}=C^{2}_{2}r^{2},
\end{equation}

\begin{equation}
e^{\lambda}=\left(\frac{C_3}{\psi}\right)^{2},
\end{equation}

\begin{equation}
\xi^{i}=C_1\delta_{n+2}^{i}+\left( \frac{\psi
r}{2}\right)\delta_1^{i},
\end{equation}
Where $C_2$ and $C_3$ are constants of integrations.

Equations (15)-(17) help us to write Einstein-Maxwell field equations 
(5)-(9) in the following form
\begin{equation}
\frac{n(n-1)}{2r^{2}}\left(1-\frac{\psi^{2}}{C_3^{2}}\right)-\frac{n\psi
\psi'}{rC_3^{2}}=8\pi \rho+E^{2},
\end{equation}

\begin{equation}
\frac{n}{2r^{2}}\left[(n+1)\frac{\psi^{2}}{C_3^{2}}-(n-1)\right]=8\pi
p-E^{2},
\end{equation}

\begin{equation}
\frac{n\psi\psi'}{rC_3^{2}}+n(n-1)\frac{\psi^{2}}{2r^{2}C_3^{2}}-\frac{(n-1)(n-2)}{2r^{2}}=8\pi
p +E^{2}.
\end{equation}

One can note that in Eqs. (18)-(20) we have four unknowns
($\rho$,~$p$,~$\psi$,~$E$) with three equations. So to solve the
above equations let us assume that the pressure is linearly
dependent on the density, i.e.
\begin{equation}
p=\alpha \rho +\beta,
\end{equation}
where $0<\alpha <1 $ and $\beta$ is some arbitrary constant. 
It is to note that $\alpha$ here is a constant which has a 
relation with the sound speed as $dp_r/d\rho = \alpha$, and 
$\beta$ is arbitrary constant that is related to the dimension of the spacetime.
 
Solving Eqs. (18)-(20) with the help of equation (21), one can
obtain
\begin{equation}
\psi^{2}=\frac{(n-1)^{2}}{n}\frac{(\alpha+1)C_3^{2}}{n(1+\alpha)-2\alpha}+\frac{16\pi\beta
C_3^{2}}{(1+\alpha)(n+1)}r^{2}+C_4r^{-\frac{2(n+n\alpha-2\alpha)}{1+3\alpha}},
\end{equation}
where $C_4$ is a constant of integration. Let us assume $C_4=0$ to
avoid the infinite mass at the origin. Now $e^{-\lambda}$ can be
obtained as
\begin{equation}
e^{-\lambda}=\frac{\psi^{2}}{C_3^{2}}=\frac{(n-1)^{2}}{n}\frac{(\alpha+1)}{n(1+\alpha)-2\alpha}+\frac{16\pi\beta}{(1+\alpha)(n+1)}r^{2}.
\end{equation}

The matter density and isotropic pressure can be obtained as
\begin{equation}
\rho=\frac{1}{n(1+\alpha)-2\alpha}\frac{(n-1)^{2}}{8\pi
r^{2}}-\frac{\beta n}{1+\alpha},
\end{equation}

\begin{equation}
p=\frac{\alpha}{n(1+\alpha)-2\alpha}\frac{(n-1)^{2}}{8\pi
r^{2}}+\beta\left(1-\frac{\alpha n}{1+\alpha}\right).
\end{equation}

The expression of electric field becomes
\begin{equation}
E^{2}=\frac{(n-1)(1-\alpha)}{n(1+\alpha)-2\alpha}~\frac{1}{2r^{2}}=\frac{q}{r^{n-2}}.
\end{equation}

\begin{figure*}[thbp]
\begin{center}
\vspace{0.5cm}
\includegraphics[width=0.45\textwidth]{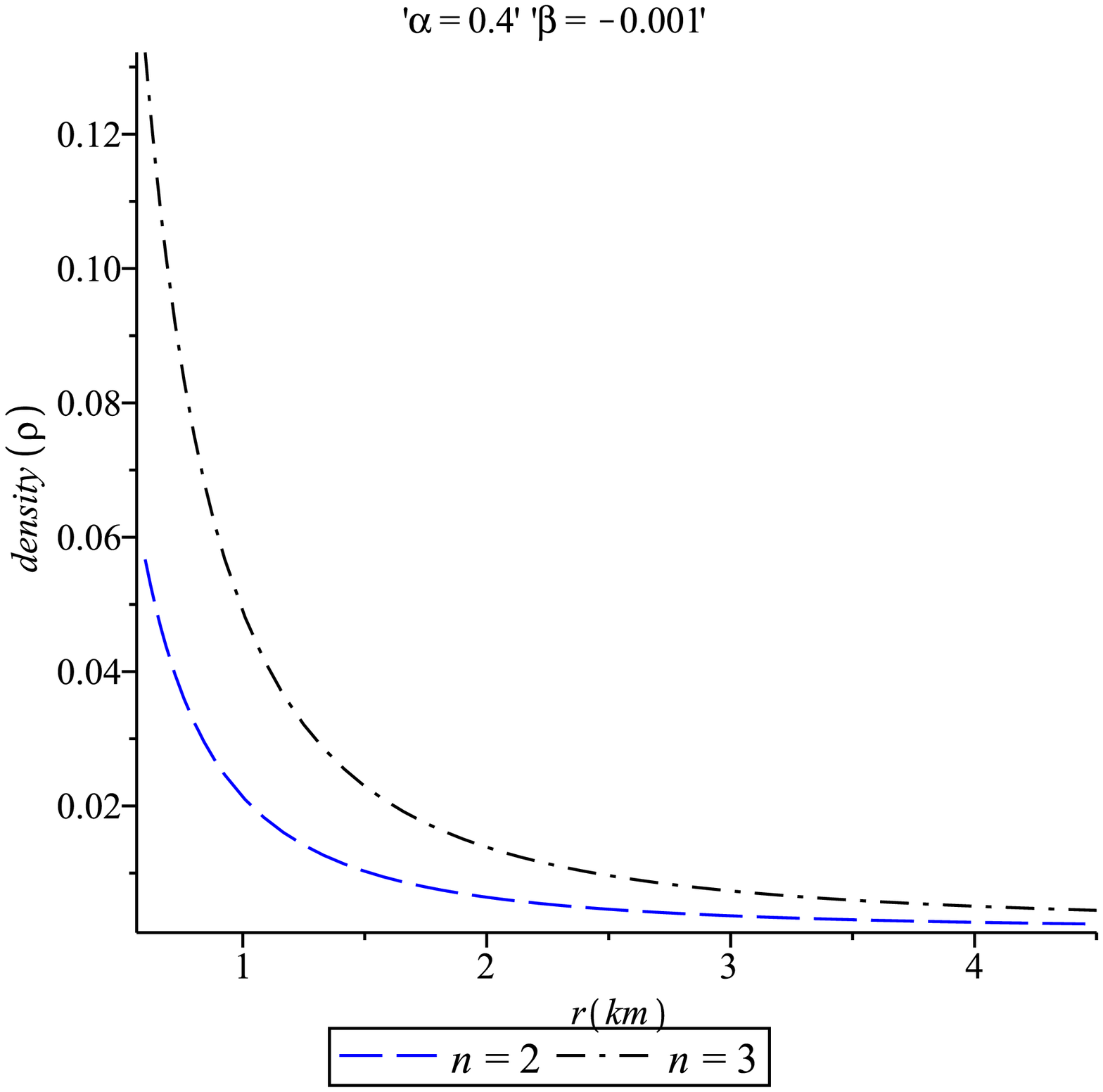}
\includegraphics[width=0.45\textwidth]{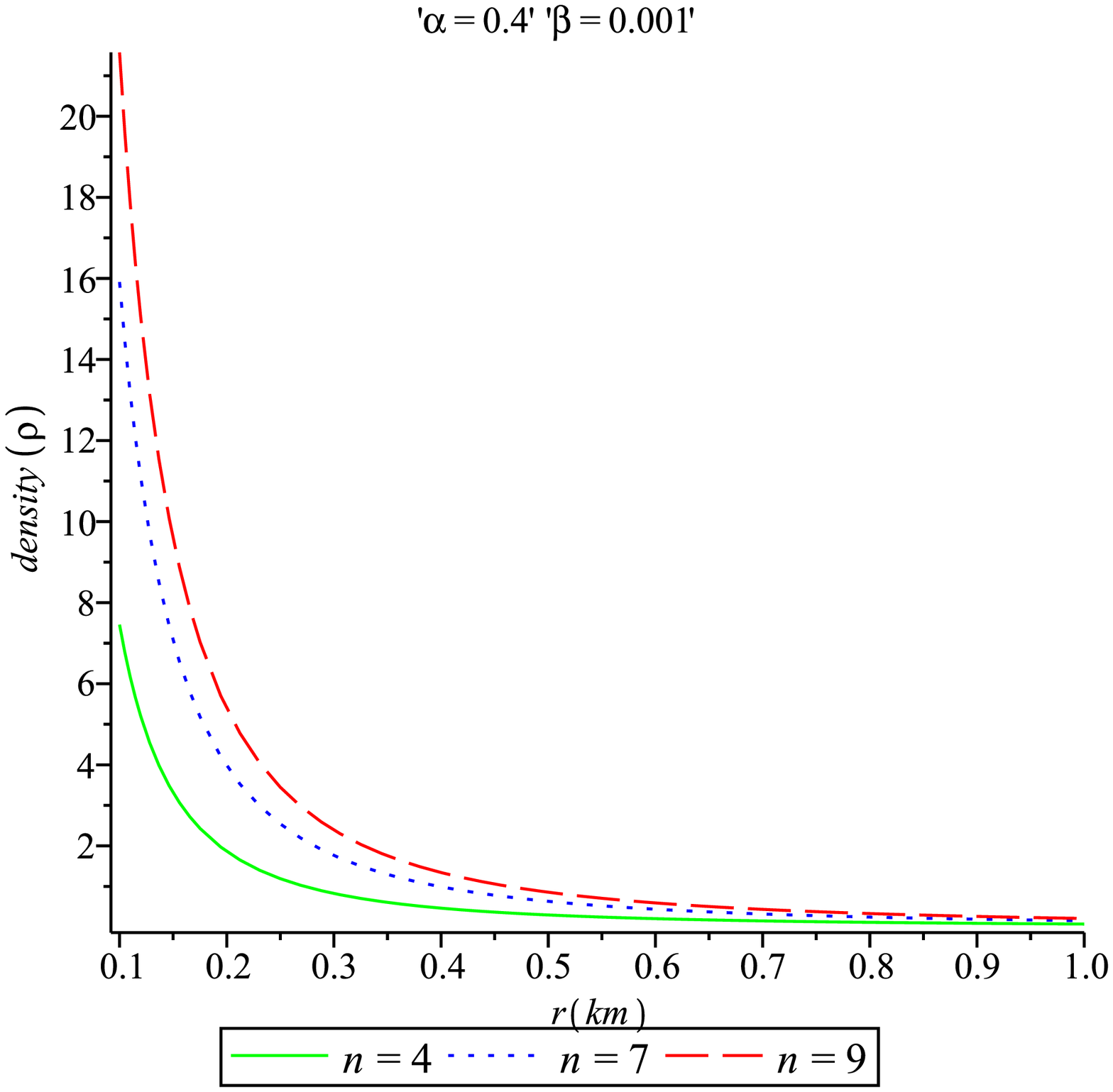}
\caption{Variation of  density $\rho$ against $r$ in the interior
of the compact star for different dimensions: $4D$ and $5D$ in the left panel whereas $6D$, $9D$ and $11D$ in the right panel}
\end{center}
\end{figure*}

\subsection{The exterior spacetime and matching condition} The
solution of the Einstein-Maxwell equation in higher dimensional
spacetime for $r>R$ is given by the following
Reissner-Nordstr\"{o}m (RN) spacetime in higher dimension as
\begin{equation}
ds^{2}=-\left(1-\frac{\mu}{r^{n-1}}+\frac{Q^{2}}{r^{2(n-1)}}\right)dt^{2}+
\left(1-\frac{\mu}{r^{n-1}}+\frac{Q^{2}}{r^{2(n-1)}}\right)^{-1}dr^{2}+r^{2}d\Omega^{2},
\end{equation}
where $\mu$ is related to the mass $M$ as $\mu=\frac{16\pi G M}{n
{\Omega}_n}$ and $Q$ is its charge. So our interior solution
should match with Eq. (31) at the boundary $r=R$.

The continuity of the metric $e^{\nu}$ gives the constant $C_2$ as
\begin{equation}
C_2=\sqrt{\frac{1}{R^{2}}\left[1-\frac{\mu}{R^{n-1}}+\frac{Q^{2}}{R^{2(n-1)}}
\right]}
\end{equation}
and the continuity of $e^{\lambda}$ gives
\begin{equation}
Q^{2}=R^{2(n-1)}\left[\frac{(n-1)^{2}}{n}\frac{\alpha+1}{n+n\alpha-2\alpha}+\frac{16\pi\beta
R^{2}}{(\alpha+1)(n+1)}-1+\frac{\mu}{R^{n-1}} \right].
\end{equation}

The intensity of the electric field at the boundary can be
obtained as
\begin{equation}
E(R)=\frac{Q(R)}{R^{n-2}}=R\sqrt{\frac{(n-1)^{2}}{n}\frac{\alpha+1}{n+n\alpha-2\alpha}+\frac{16\pi\beta
R^{2}}{(\alpha+1)(n+1)}-1+\frac{\mu}{R^{n-1}}}.
\end{equation}

The profile of the electric field is shown in Fig. 3. The figure
shows that $E^{2}$ is a monotonic decreasing function of $r$ which
attains maximum value for $11D$ and minimum value for $4D$.

\subsection{The Junction Condition}

Here the metric coefficients are continuous at $r=R$, but that does not 
ensure that their derivatives are also continuous at the junction surface. 
In other words the affine connections may be discontinuous at the 
junction surface $r=R$. To take care of this we use the Darmois-Israel 
formation to determine the surface stresses at the junction boundary. 
The intrinsic surface stress energy tensor as given by Lancozs equations is 
as follows:

Obviously the metric coefficients are continuous at but it
does not ensure that their derivatives are also continuous at the
junction surface. In other words the affine connections may be
discontinuous there. To take care of this let us use the
Darmois-Israel formation to determine the surface stresses at the
junction boundary. The intrinsic surface stress energy tensor
$S_{ij}$ is given by Lancozs equations in the following form
\begin{equation}
S^{i}_{j}=-\frac{1}{8\pi}(\kappa^{i}_j-\delta^{i}_j\kappa^{k}_k).
\end{equation}

The discontinuity in the second fundamental form is given by
\begin{equation}
\kappa_{ij}=K_{ij}^{+}-K_{ij}^{-},
\end{equation}
where the second fundamental form is given by
\begin{equation}
K_{ij}^{\pm}=-n_{\nu}^{\pm}\left[\frac{\partial^{2}X_{\nu}}{\partial
\xi^{i}\partial\xi^{j}}+\Gamma_{\alpha\beta}^{\nu} \frac{\partial
X^{\alpha}}{\partial \xi^{i}}\frac{\partial X^{\beta}}{\partial
\xi^{j}} \right]|_S.
\end{equation}

Here $n_{\nu}^{\pm}$ are the unit normal vector defined by
\begin{equation}
n_{\nu}^{\pm}=\pm\left|g^{\alpha\beta}\frac{\partial f}{\partial
X^{\alpha}}\frac{\partial f}{\partial X^{\beta}}
\right|^{-\frac{1}{2}}\frac{\partial f}{\partial X^{\nu}},
\end{equation}
with $n^{\nu}n_{\nu}=1$. Here $\xi^{i}$ is the intrinsic
coordinate on the shell. $+$ and $-$ corresponds to exterior i.e,
RN spacetime in higher dimension and interior (our) spacetime
respectively.

Considering the spherical symmetry of the spacetime surface stress
energy tensor can be written as
$S^{i}_j=diag(-\Sigma,\mathcal{P},\mathcal{P},...,\mathcal{P})$
where $\Sigma$ and $\mathcal{P}$ are the surface energy density
and surface pressure respectively and can be provided by
\begin{equation}
\Sigma=-\frac{n}{4\pi
R}\sqrt{1-\frac{\mu}{R^{n-1}}+\frac{q^{2}}{R^{2(n-1)}}}
+\frac{n}{4\pi
R}\sqrt{\frac{(n-1)^{2}}{n}\frac{\alpha+1}{n+n\alpha-2\alpha}
+\frac{16\pi \beta}{(1+\alpha)(n+1)}r^{2}},
\end{equation}

\begin{eqnarray}
\mathcal{P}=\frac{n-1}{4\pi R}\left(1+\frac{\mu}{2R^{n-1}}-\frac{
q^{2}}{R^{2n-1}} \right)\sqrt{1+\frac{\mu}{2R^{n-1}}-\frac{
q^{2}}{R^{2n-2}}} \nonumber \\ -\frac{n}{4\pi
R}\sqrt{\frac{(n-1)^{2}}{n}\frac{\alpha+1}{n+n\alpha-2\alpha}+\frac{16
\pi \beta R^{2}}{(\alpha+1)(n+1)}}.
\end{eqnarray}

The mass of the thin shell can be obtained as
\begin{equation}
m_s=\frac{2\pi^{\frac{n+1}{2}}}{\Gamma\left(\frac{n+1}{2}\right)}
R^{n}\Sigma.
\end{equation}

Now using Eqs. (40) and (41) one can obtain the mass of the
charged fluid sphere in terms of the mass of the thin shell as
\begin{equation}
\mu=R^{n-1}\left[1+\frac{q^{2}}{R^{2(n-1)}}-G^{2}+2BG-B^{2}\right],
\end{equation}
where
\[G=\frac{2m_s\Gamma\left(\frac{n+1}{2}\right)}{nR^{n-1}\pi^{\frac{n-1}{2}}}\]
and
\[B=\sqrt{\frac{(n-1)^{2}}{n}\frac{\alpha+1}{n+n\alpha-2\alpha}+\frac{16
\pi \beta R^{2}}{(\alpha+1)(n+1)}}.\]

\section{Physical Analysis of the solutions}
For a physically meaningful solution one must have pressure and
density are decreasing function of $r$. For our model
\begin{equation}
\frac{dp}{dr}=-\frac{(n-1)^{2}}{n+\alpha(n-2)}\frac{\alpha}{4\pi
r^{3}}<0,
\end{equation}

\begin{equation}
\frac{d\rho}{dr}=-\frac{(n-1)^{2}}{n+\alpha(n-2)}\frac{1}{4\pi
r^{3}}<0.
\end{equation}

The above expression indicates that both $\rho$ and $p$ are
monotonic decreasing function of $r$, i.e. the have maximum value
at the center of the star and it decreases radially outwards. The
constant of integration $\beta$ can be obtained by imposing the
condition $p(r=R)=0$ as
\begin{equation}
\beta=\frac{\alpha(\alpha+1)(n-1)^{2}}{8\pi
R^{2}(n+n\alpha-2\alpha)(n\alpha-\alpha-1)}.
\end{equation}

The above equations consequently gives
\[R_n=\sqrt{\frac{\alpha(\alpha+1)(n-1)^{2}}{8\pi \beta (n+n\alpha-2\alpha)(n\alpha-\alpha-1)}},\] 
where $n=D-2$,~$D$ being the dimension of the spacetime. 
One can easily verify that \[\frac{dp}{d\rho}=\alpha.\]

To satisfy the causality conditions one must have $0<
\frac{dp}{d\rho} <1$ which implies $0<\alpha<1$. To find the
radius of different dimensional charged star let us fix
$\alpha=0.4$. From the expression of $R_n$ one can note that for
$\alpha=0.4$,~$n+(n-2)\alpha$ is always positive and
$n\alpha-\alpha-1$ is negative for $n=2$ and $n=3$, i.e. for four
and five dimensional spacetime. So we must have $\beta<0$ for
$n=2$ and $n=3$. On the other hand $n\alpha-\alpha-1$ is positive
for six dimensional onwards when $\alpha=0.4$. So we have to take
positive beta for six dimensional onwards. So to find the radius
of the charged star in different dimension we have choose
$\alpha=0.4, \beta= -0.001 $ for $4D$ and $5D$ spacetime and
$\alpha=0.4, \beta= 0.001 $ for the spacetime onwards six
dimension. The radius of the charged star in different dimension
are shown in Table 1. From Fig. 2, we see that the radius of the
star is found where the graphs of p(r) cut the r-axis and one can
note that the radius of the charged star in $5D$ is greater than
$4D$ for fixed values of $\alpha$ and $\beta$ mentioned in the
figure. On the other hand the radius of the charged star decreases
when the dimension increases, i.e. for for fixed values of
$\alpha$ and $\beta$ mentioned in the figure the radius is maximum
for $6D$ charged star and is minimum for $11D$ charged star.

\begin{figure}[htbp]
\centering
\includegraphics[width=0.45\textwidth]{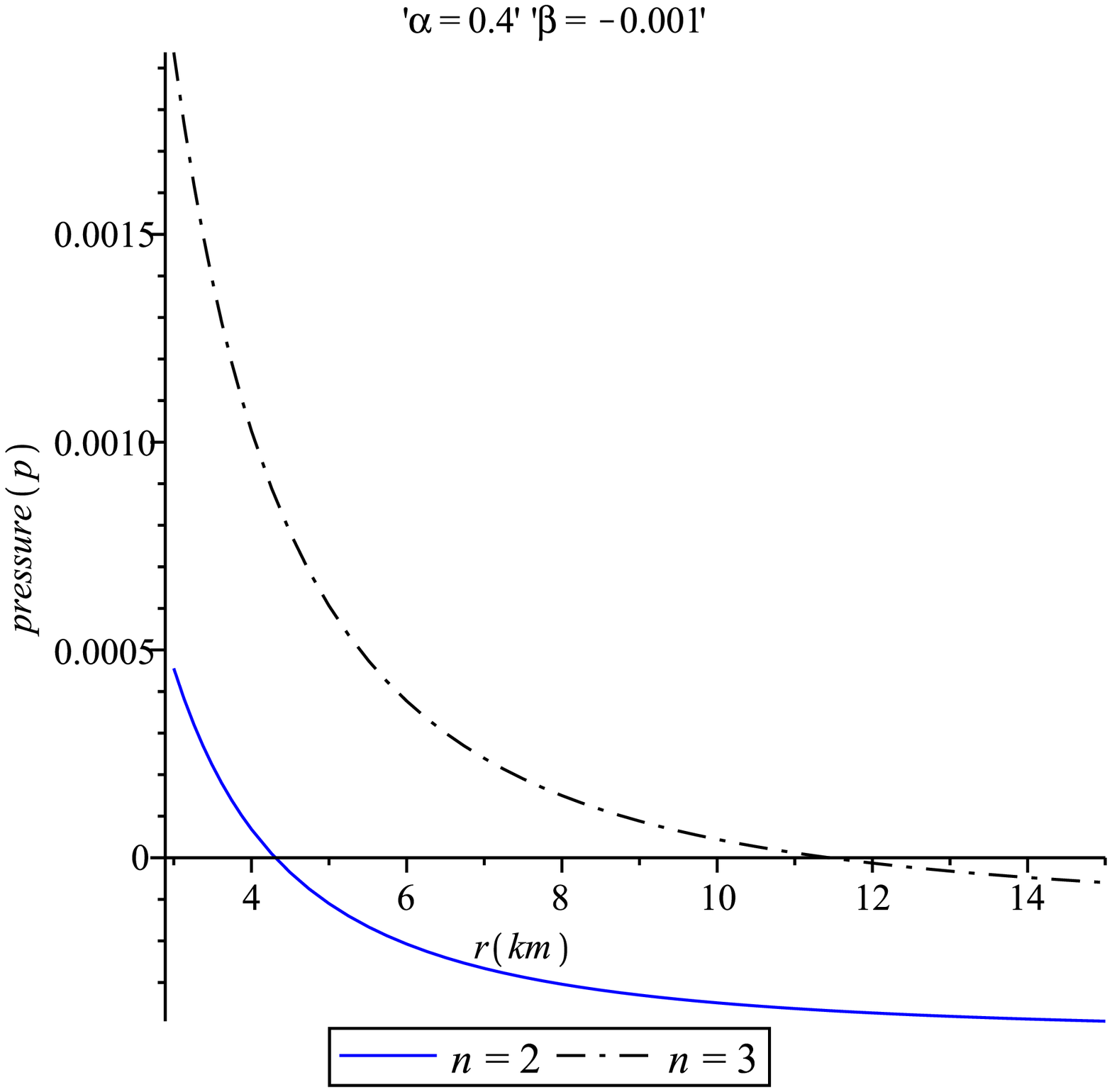}
\includegraphics[width=0.45\textwidth]{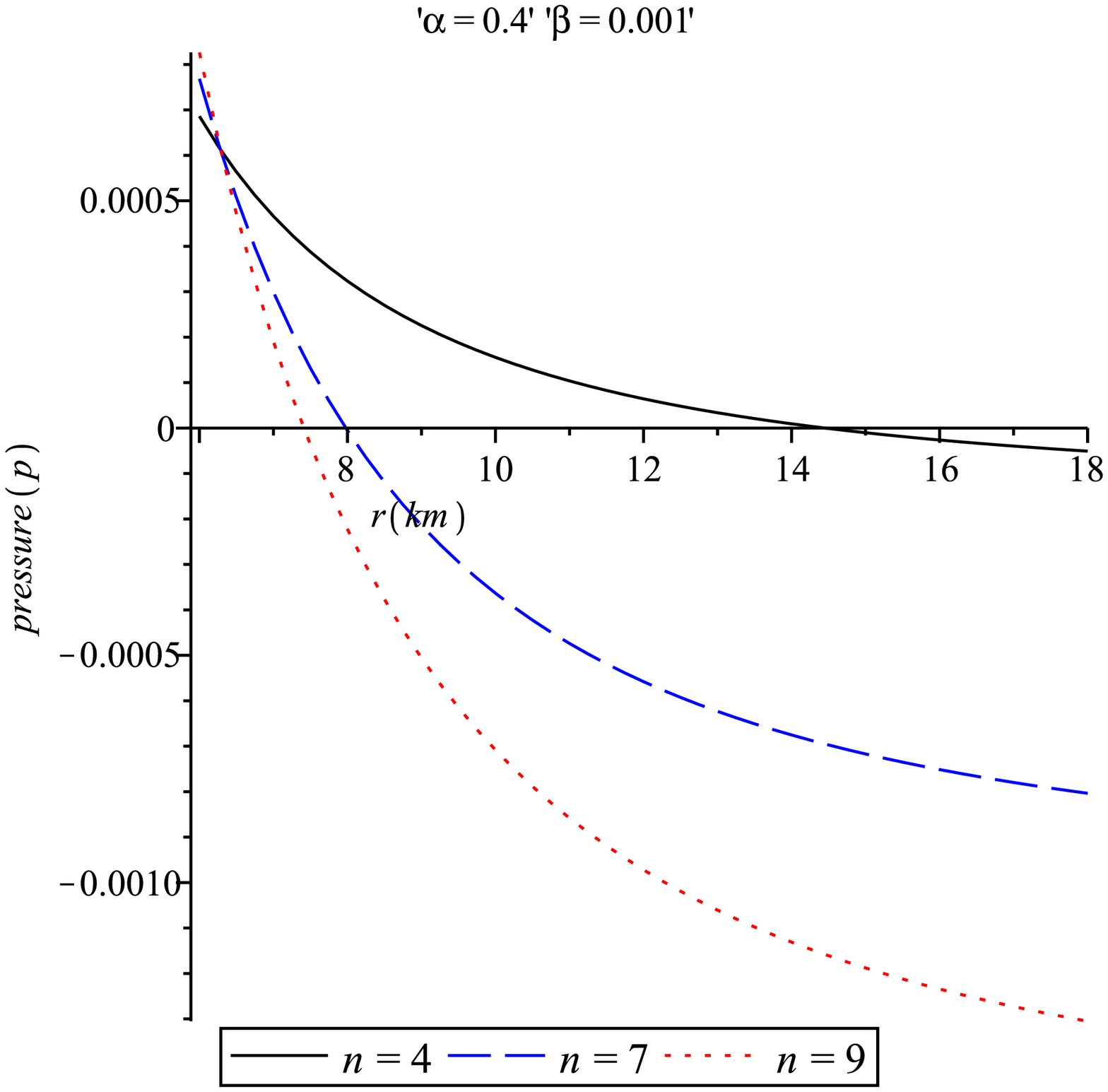}
\caption{Radii of the stars are found where radial pressures cut
$r$ axis for different dimensions: $4D$ and $5D$ in the left panel whereas $6D$,
$9D$ and $11D$ in the right panel}
\end{figure}

The gravitational mass inside the charged sphere can be obtained
as
\[m(r)=\int_0^{r}\frac{2 \pi^{\frac{n+1}{2}}}{\Gamma\left(\frac{n+1}{2}\right)}r^{n}\left[\rho+\frac{E^{2}}{8 \pi}\right]dr\]
\begin{equation}
~~~~~~~~~~~~~~~~~~~~~~~~~~~~~~~~~~~~~~~~~~~~~~~~~~~~~~~~=\frac{\pi^{\frac{n-1}{2}}}{4\Gamma\left(\frac{n+1}{2}\right)}
\left[\frac{2n-(1+\alpha)}{2(n+n\alpha-2\alpha)}r^{n-1}-\frac{8n\pi\beta}{(1+\alpha)(n+1)}r^{n+1}
\right].
\end{equation}

The profile of the mass function for different dimensional compact
stars are shown in Fig. 7. The figure indicates that $m(r)$ is a
monotonic increasing function of r and $m(r)>0$ inside the charged
fluid sphere. Moreover as $r\rightarrow 0$, $m(r)\rightarrow 0$,
i.e. the mass function is regular at the center of the charged
star.

The charged density can be obtained as
\begin{equation}
\sigma=\frac{\Gamma\left(\frac{n+1}{2}\right)}{2
\pi^{\frac{n+1}{2}}}\frac{n-1}{r^{2}}
\sqrt{\frac{(n-1)(1-\alpha)}{2(n+n\alpha-2\alpha)}}\sqrt{\frac{(n-1)^{2}}{n}\frac{\alpha+1}{n+n\alpha-2\alpha}+\frac{16\pi\beta}{(1+\alpha)(n+1)}r^{2}}.
\end{equation}

The profile of the charged density is shown in Fig. 4. The figure
indicates that it is a monotonic decreasing function of $r$ and
its values increases as dimensions increases, i.e. the value of
$\sigma$ is maximum for $11D$ and minimum for $4D$.

\begin{figure}[htbp]
\centering
\includegraphics[width=0.45\textwidth]{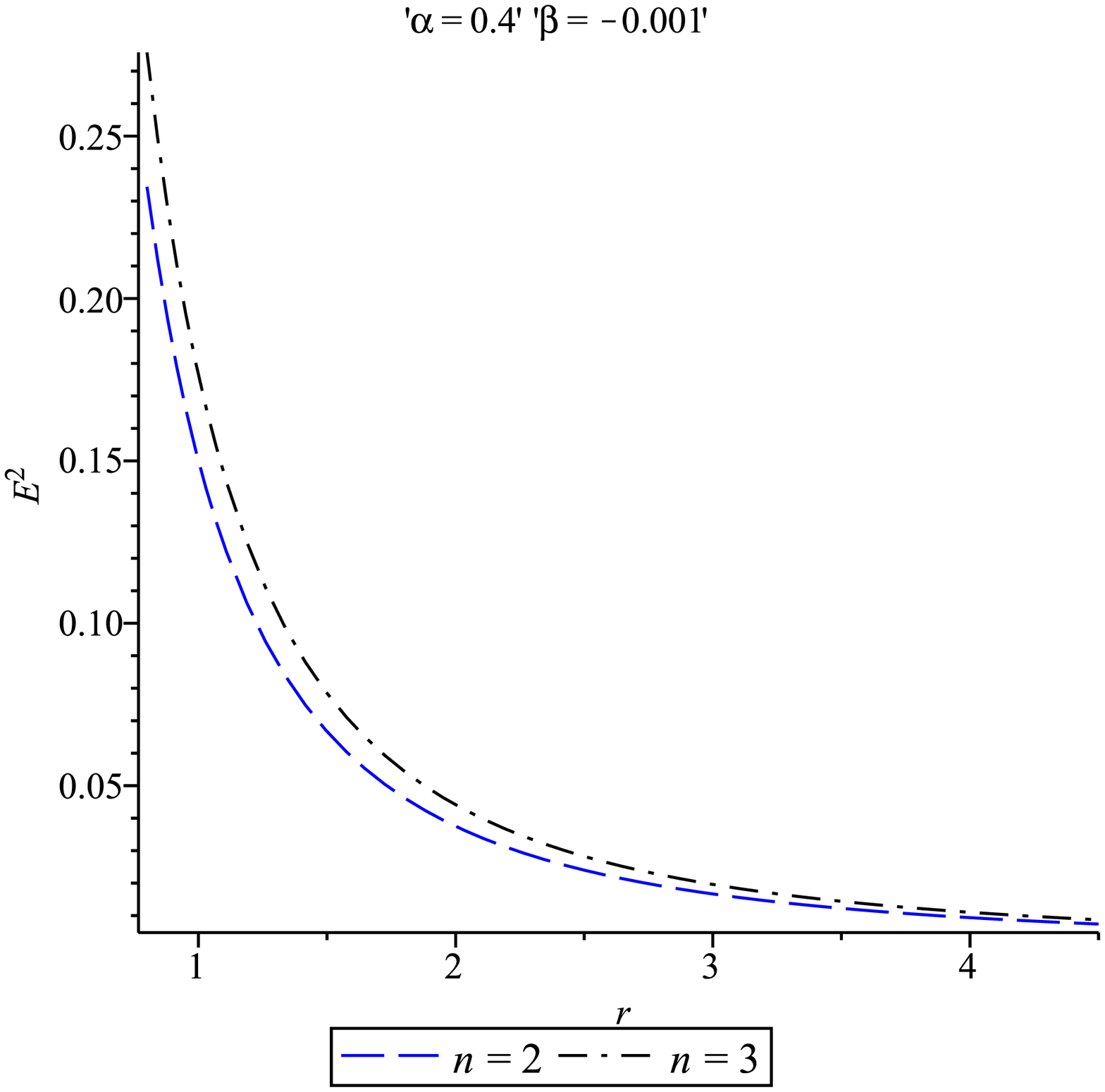}
\includegraphics[width=0.45\textwidth]{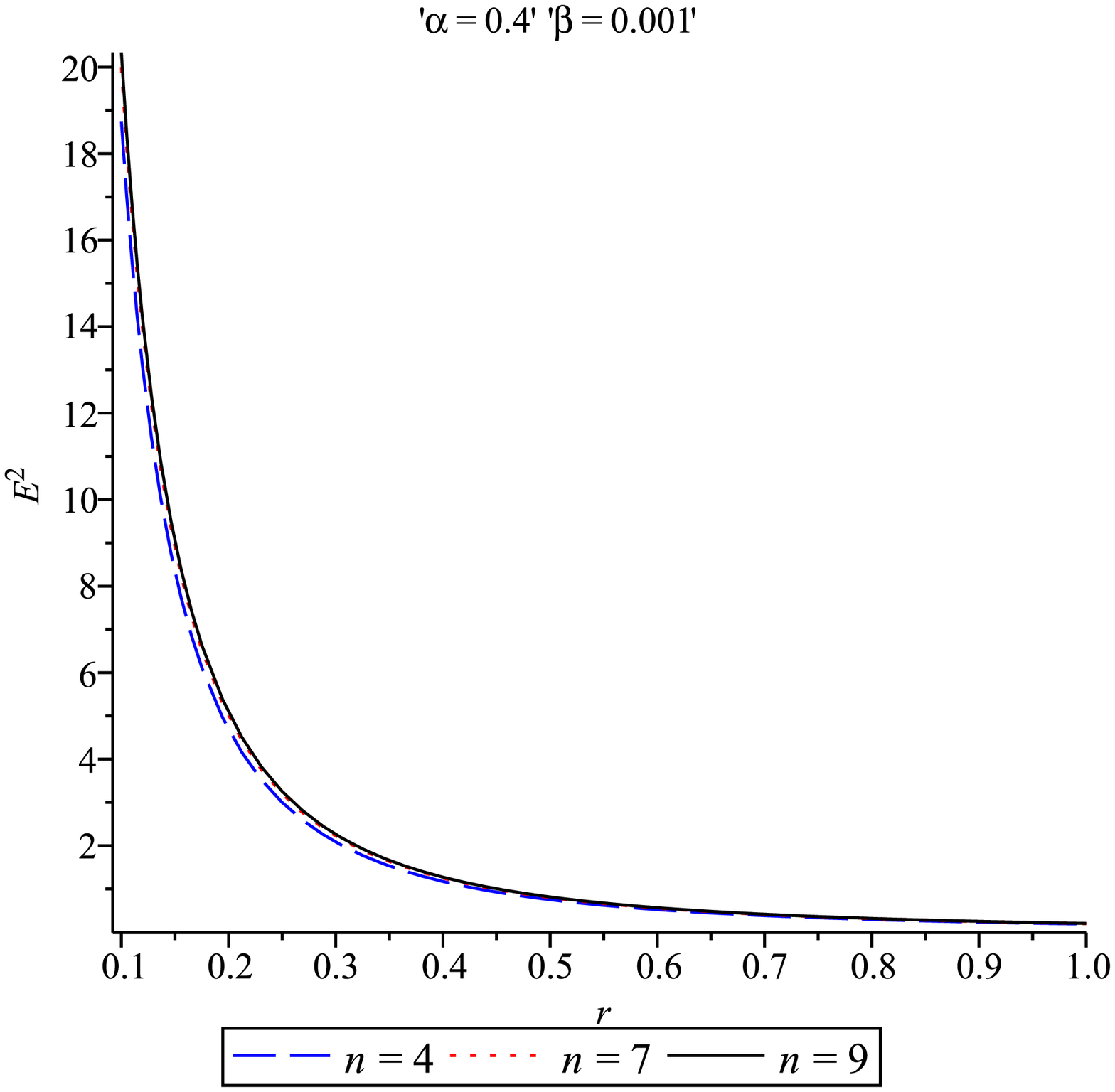}
\caption{The Electric field are plotted against $r$ for different dimensions: 
$4D$ and $5D$ in the left panel whereas $6D$, $9D$ and $11D$ in the right panel}
\end{figure}

\begin{figure}[htbp]
\centering
\includegraphics[width=0.45\textwidth]{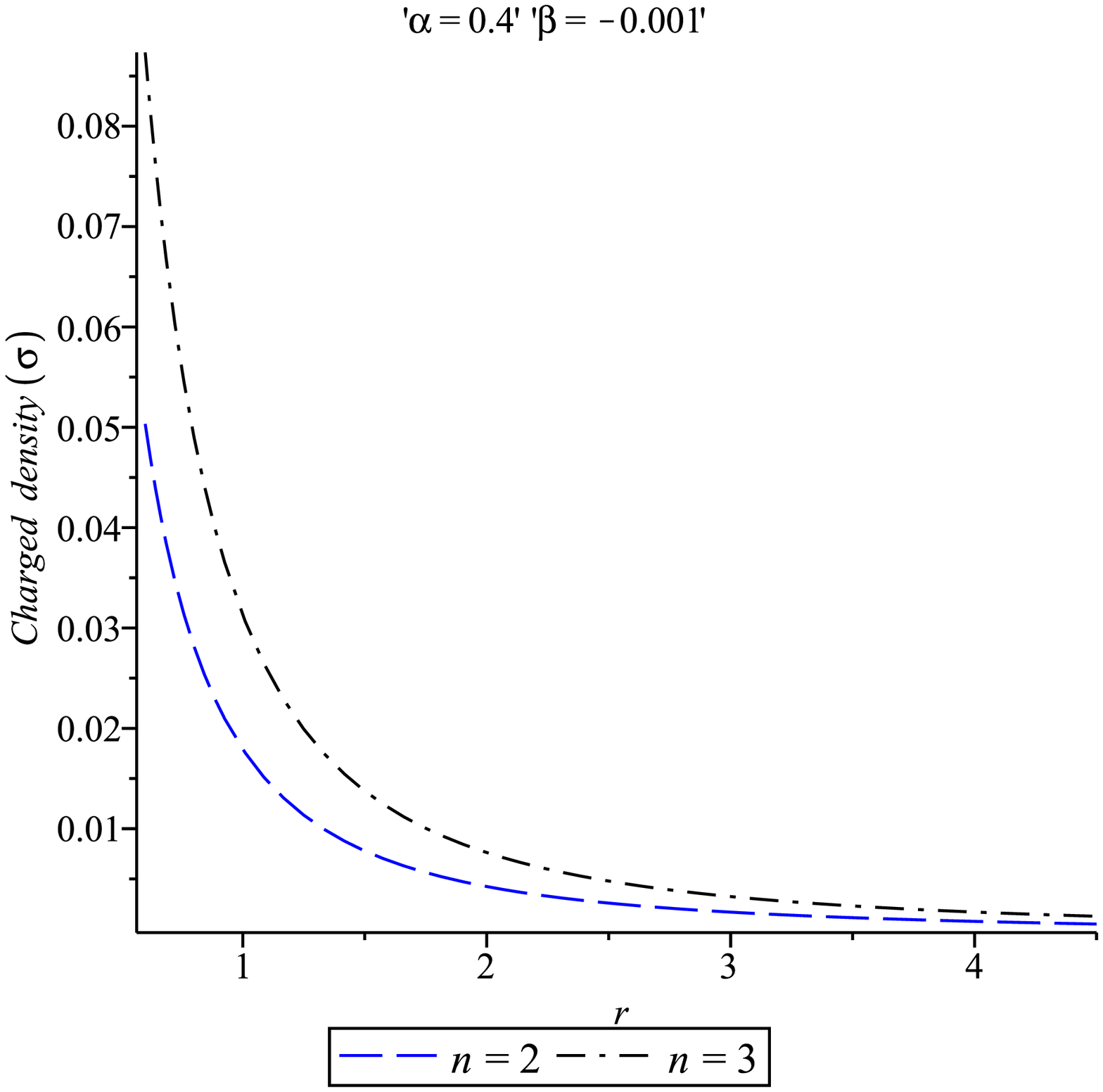}
\includegraphics[width=0.45\textwidth]{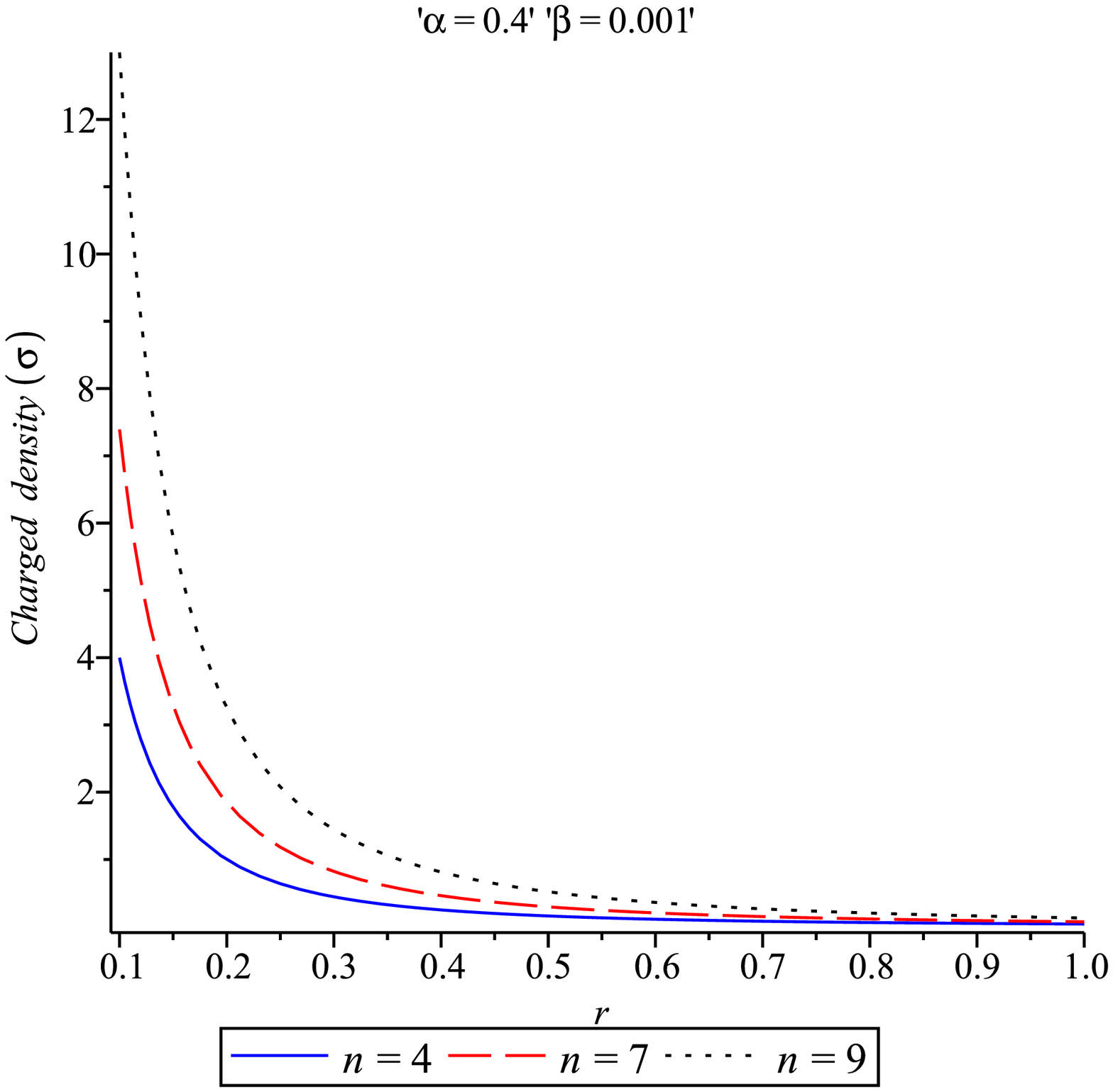}
\caption{The charged density are plotted against $r$ for different dimensions: 
$4D$ and $5D$ in the left panel whereas $6D$, $9D$ and $11D$ in the right panel}
\end{figure}

\subsection{Stability condition} To check the stability of our model under
three different forces we consider the generalized TOV equation which 
is given by the
equation~\cite{verela}
\begin{equation}
-\frac{M_G(\rho+p)}{r^{n-2}}e^{\frac{\lambda-\nu}{2}}-\frac{dp}{dr}+\sigma
\frac{q}{r^{n-2}}e^{\frac{\lambda}{2}}=0,
\end{equation}
where $M_G=M_G(r)$ is the effective gravitational mass inside a
sphere of radius $r$ which can be derived from the equation
\begin{equation}
M_G(r)=\frac{1}{2}r^{n-2}e^{\frac{\nu-\lambda}{2}}\nu'
\end{equation}
named as Tolman-Whittaker formula. The above equation describes
the equilibrium condition of the fluid sphere subject to
gravitational,hydrostatics and electric forces. Eq. (44) can be
modified in the form
\begin{equation}
F_g+F_h+F_e=0,
\end{equation}
where
\begin{equation}
F_g=-\frac{\nu'}{2}(\rho+p)=-\frac{1}{r}\left[\frac{\alpha+1}{n(1+\alpha)-2\alpha}\frac{(n-1)^{2}}{8\pi
r^{2}}-\beta(n-1)\right],
\end{equation}

\begin{equation}
F_h=-\frac{dp}{dr}=\frac{(n-1)^{2}}{n+\alpha(n-2)}\frac{\alpha}{4\pi
r^{3}},
\end{equation}

\begin{equation}
F_e=\sigma \frac{q}{r^{n-2}}e^{\frac{\lambda}{2}}.
\end{equation}

The profiles of $F_g,~F_h~F_e$ has shown in Fig. 5 for different
dimensional charged fluid sphere. The figure shows that for our
model the gravitational force $(F_g)$ is counterbalanced by the
combined effects of hydrostatic $(F_h)$ and electric forces
$(F_e)$ and thus helps to keep the model in static equilibrium.

\begin{figure}[thbp]
\centering
\includegraphics[width=0.45\textwidth]{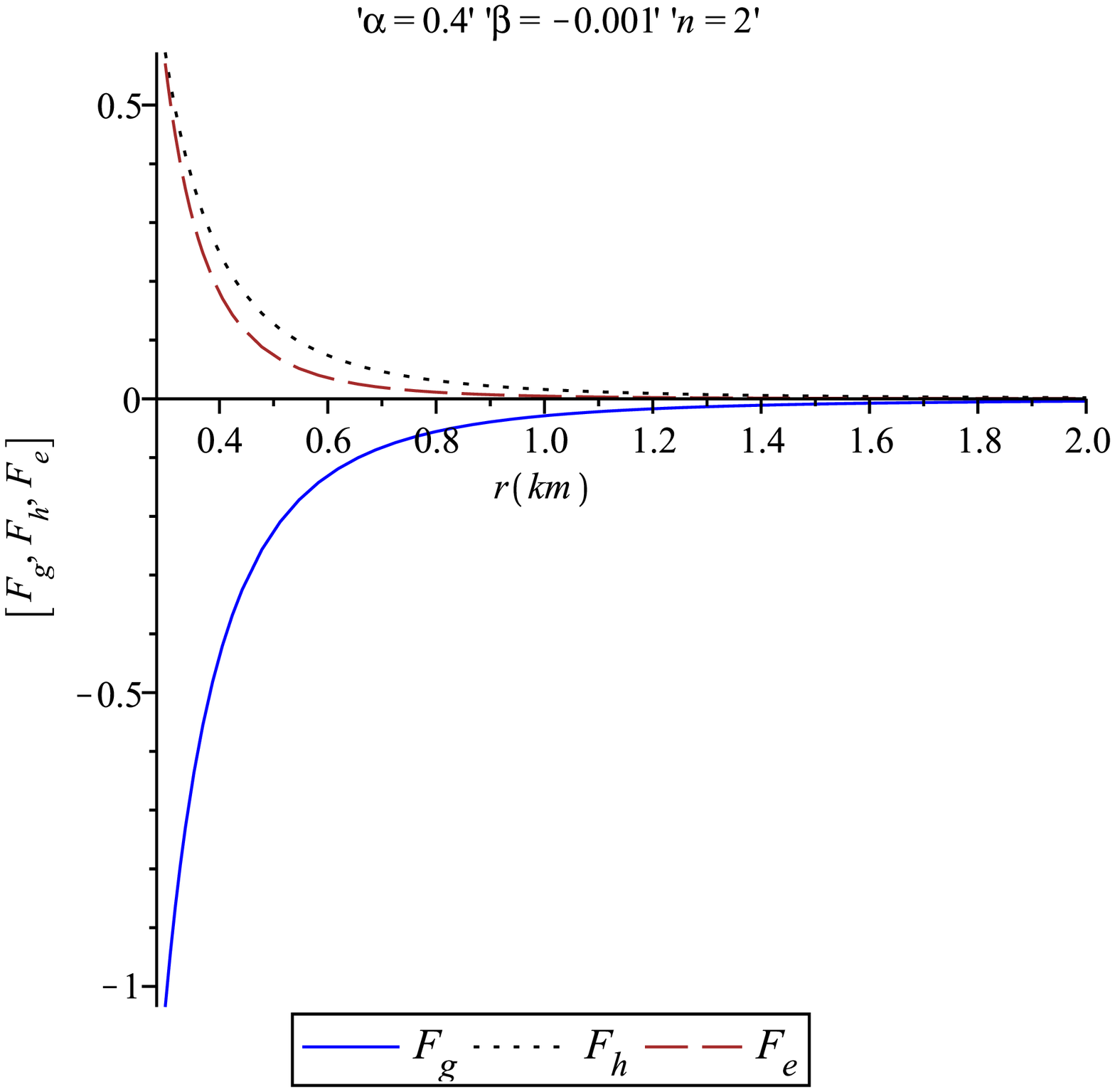}
\includegraphics[width=0.45\textwidth]{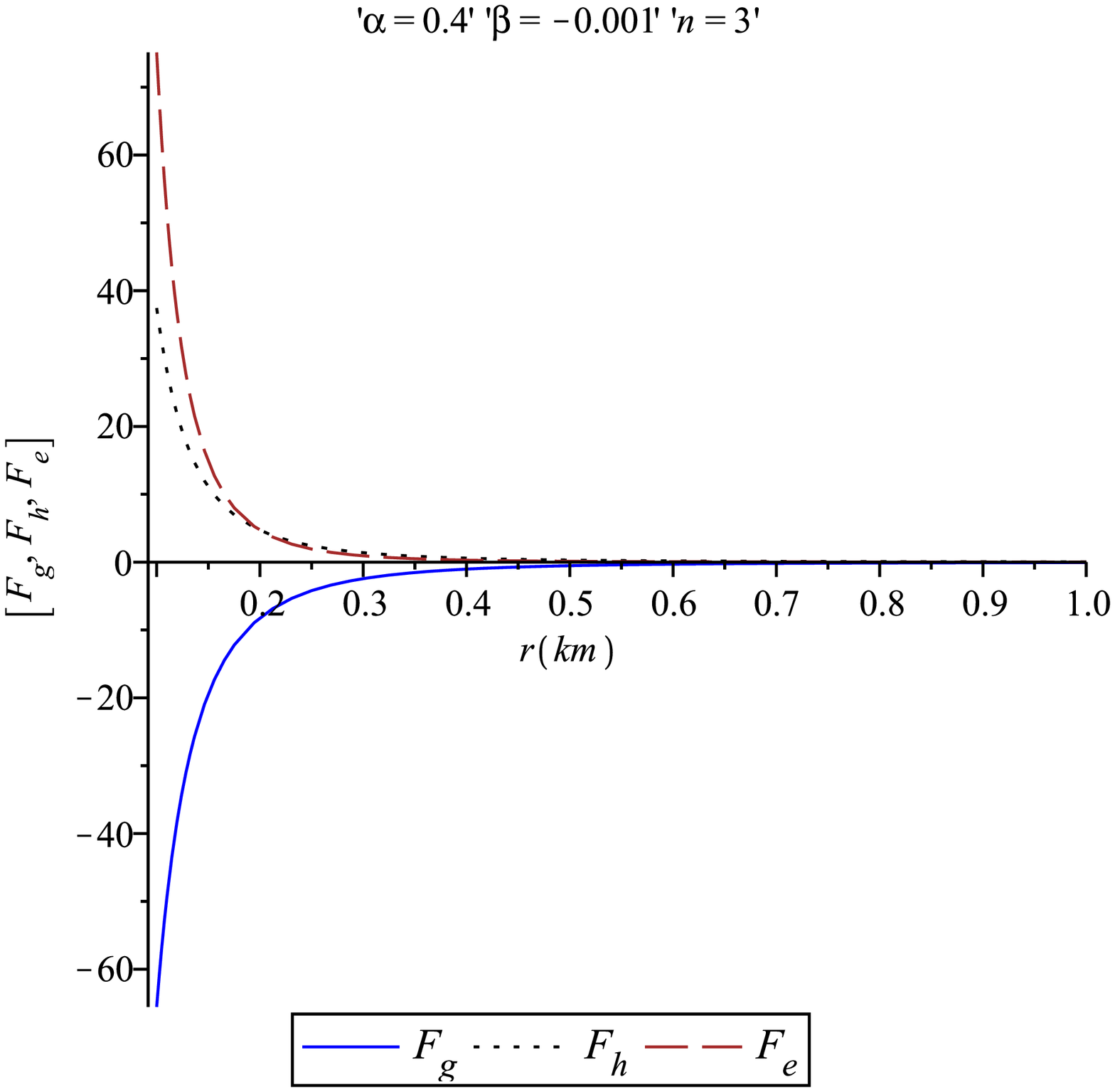}
\includegraphics[width=0.45\textwidth]{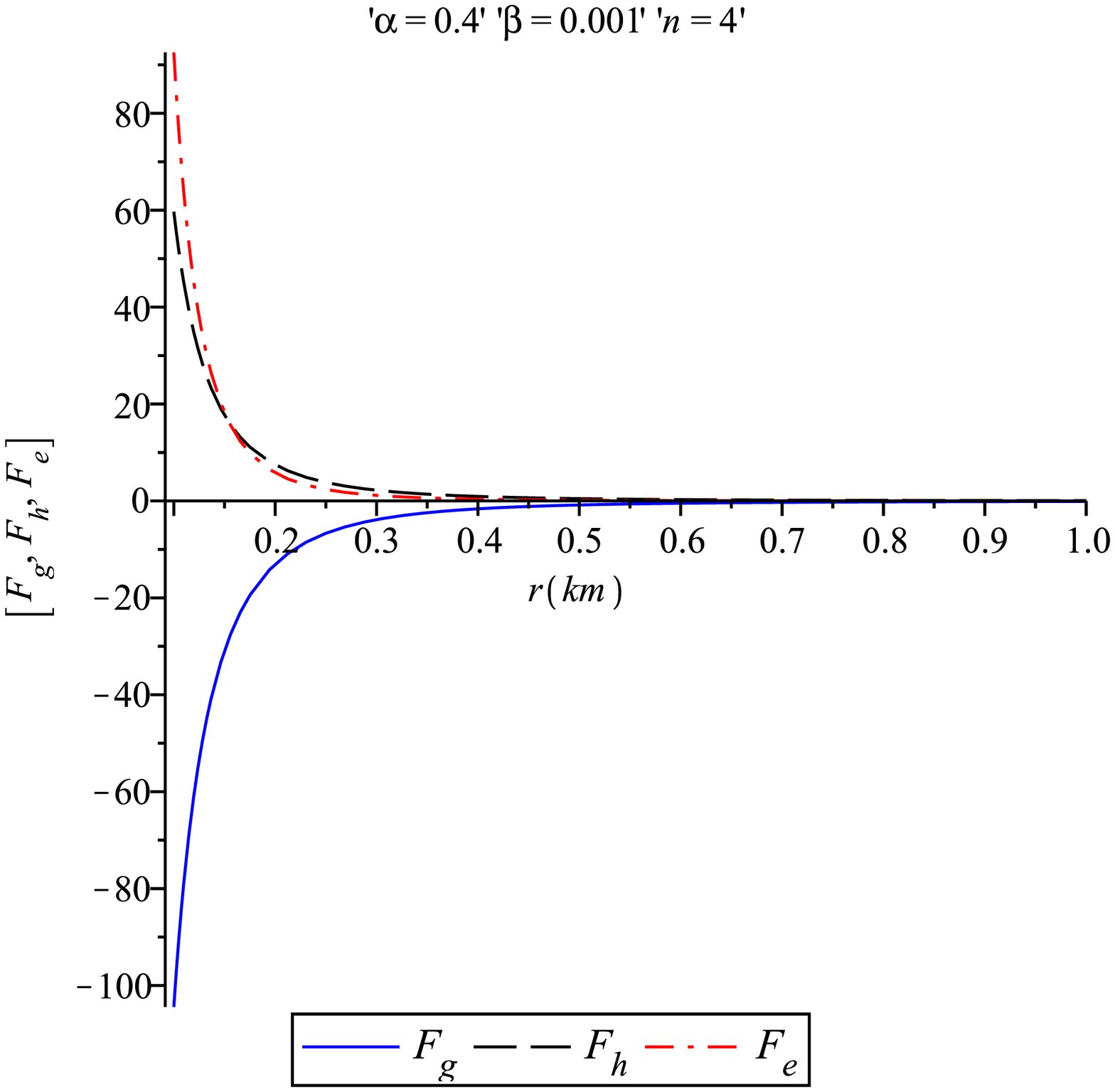}
\includegraphics[width=0.45\textwidth]{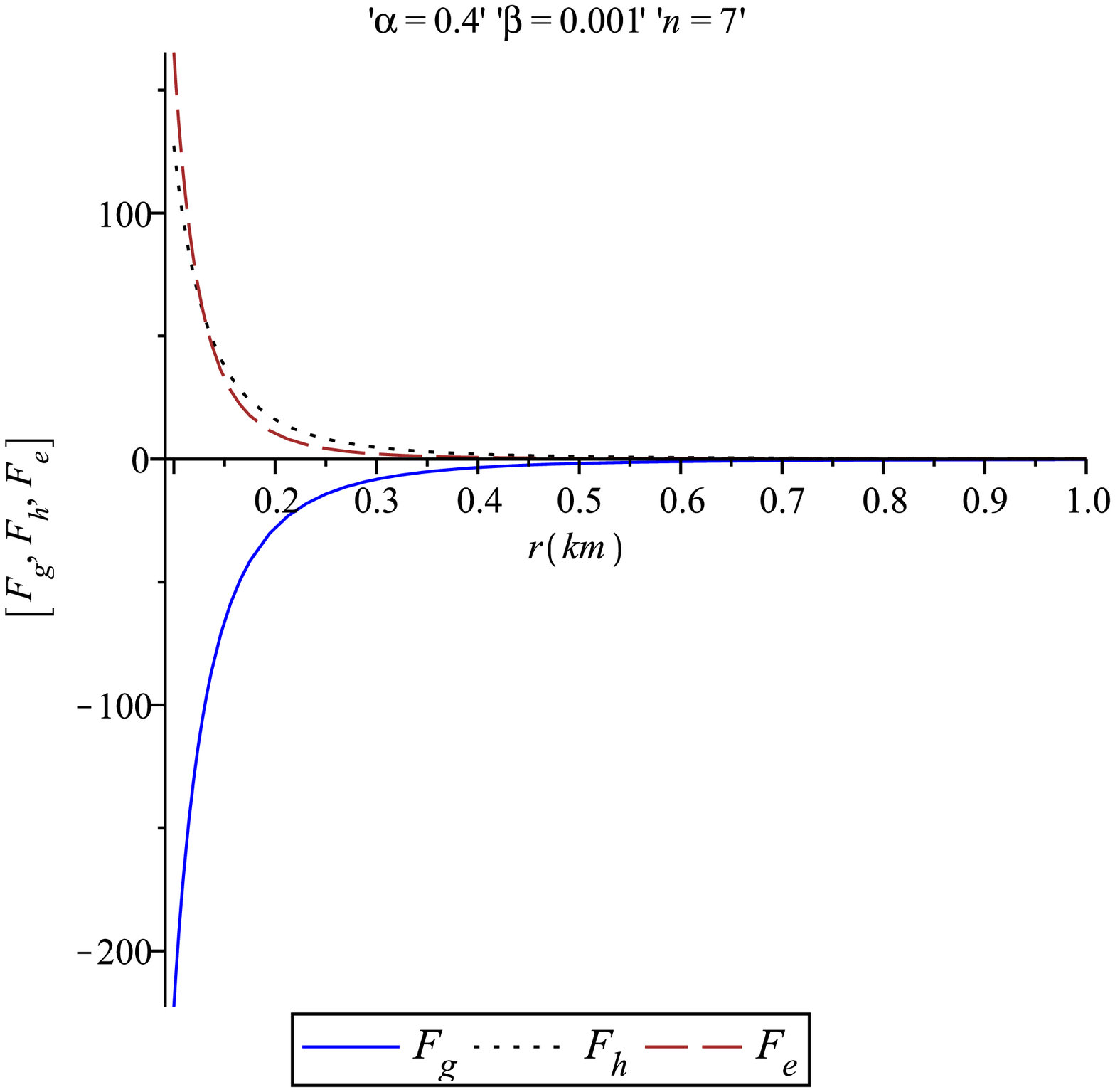}
\includegraphics[width=0.45\textwidth]{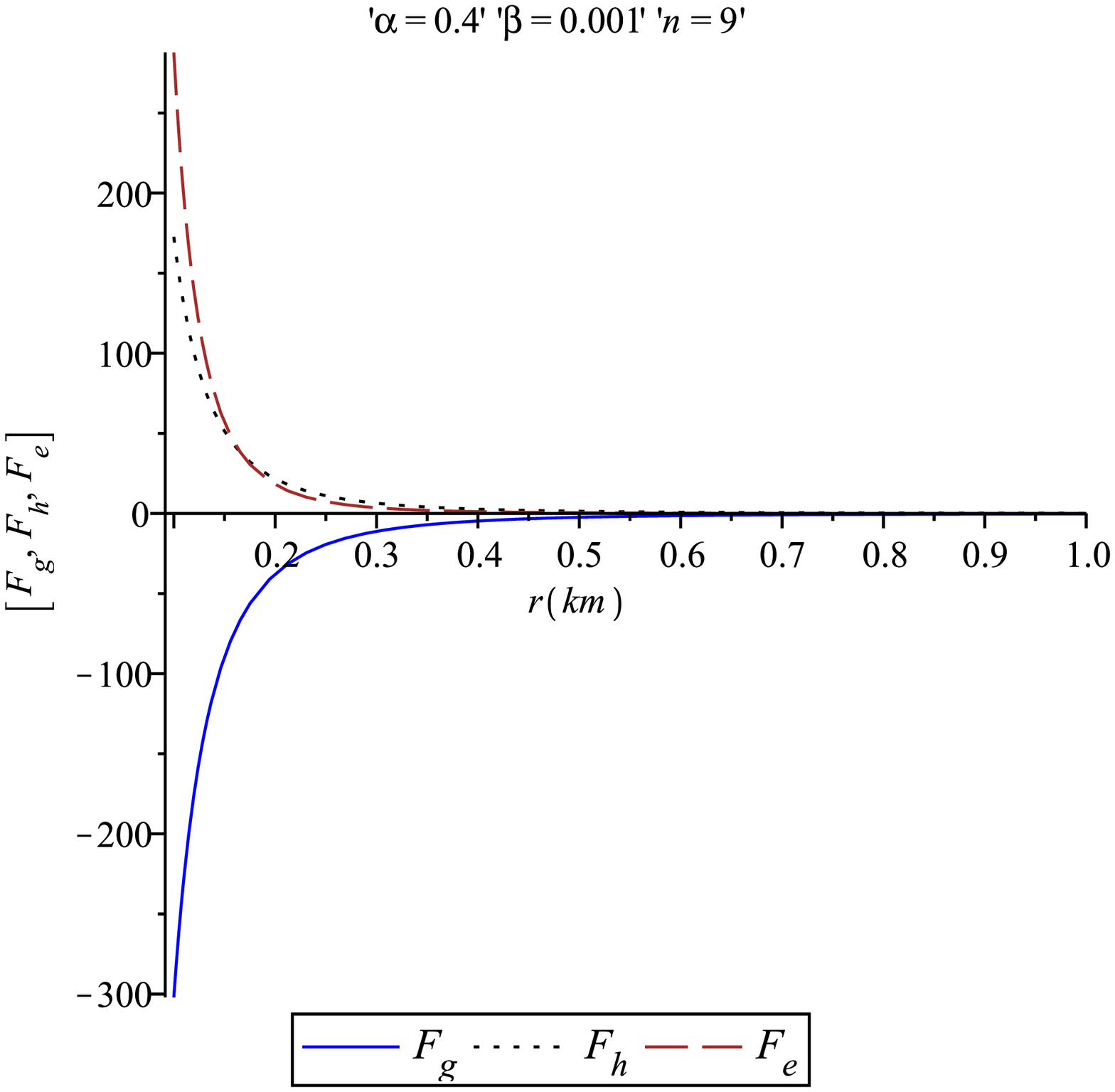}
\caption{ The three different forces, namely, gravitational forces
$(F_g)$, hydrostatic forces $(F_h)$ and electric forces $(F_e)$
are plotted against `r~(km)' for $4D$ (upper left), $5D$ (upper
right), $6D$ (middle left), $9D$ (middle right) and $11D$ (lower)}
\end{figure}

\subsection{Energy Conditions} The isotropic charged perfect fluid
sphere will satisfy the Null energy condition (NEC), Weak energy
condition (WEC) and Strong energy condition if the following
inequalities hold simultaneously inside the fluid sphere:
\begin{equation}
\rho+\frac{E^{2}}{8\pi} \geq 0,
\end{equation}

\begin{equation}
\rho+p \geq 0,
\end{equation}

\begin{equation}
\rho+p+\frac{E^{2}}{4\pi} \geq 0,
\end{equation}

\begin{equation}
\rho+3p+\frac{E^{2}}{4\pi} \geq 0.
\end{equation}

\begin{figure}[thbp]
\centering
\includegraphics[width=0.45\textwidth]{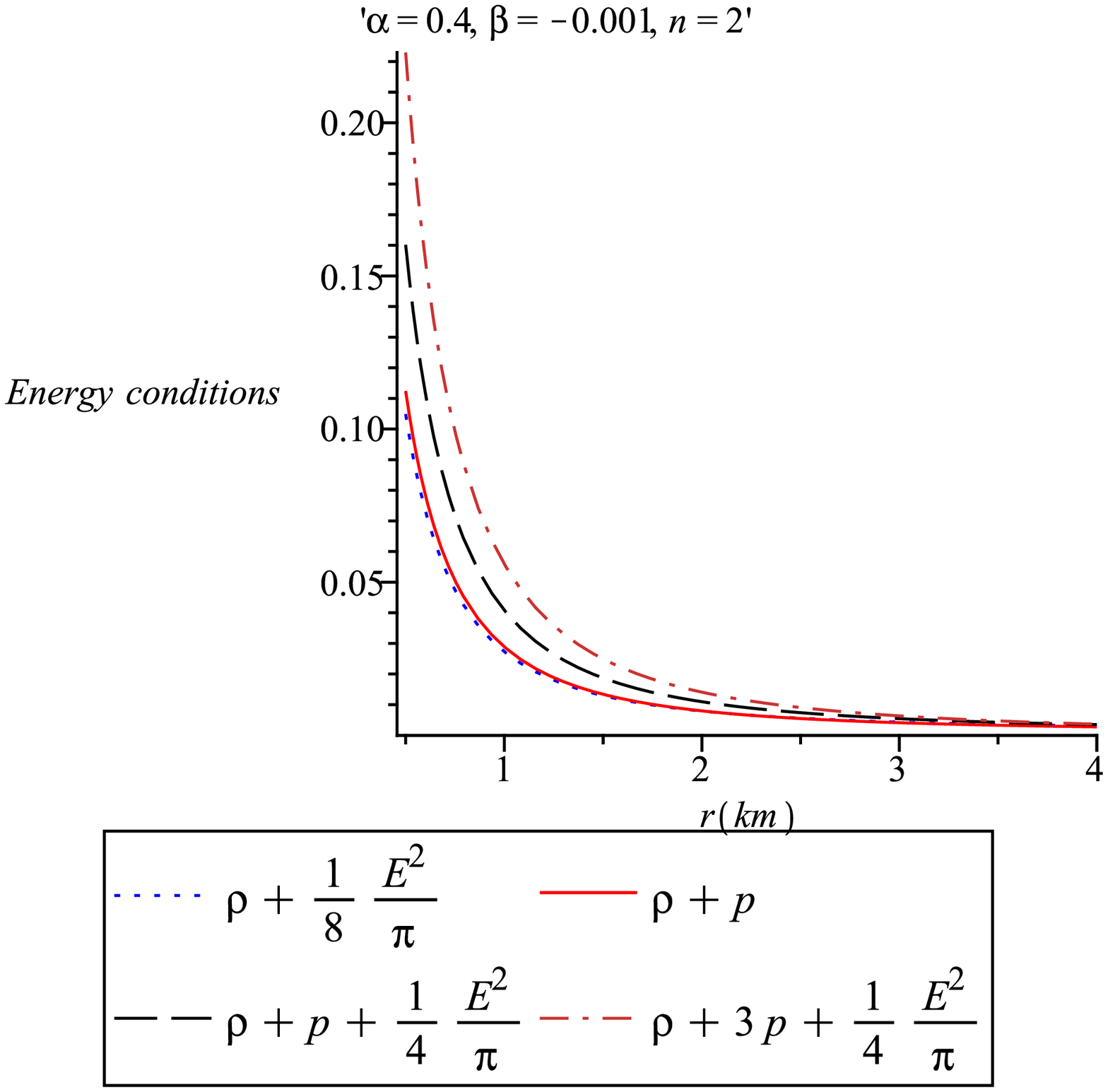}
\includegraphics[width=0.45\textwidth]{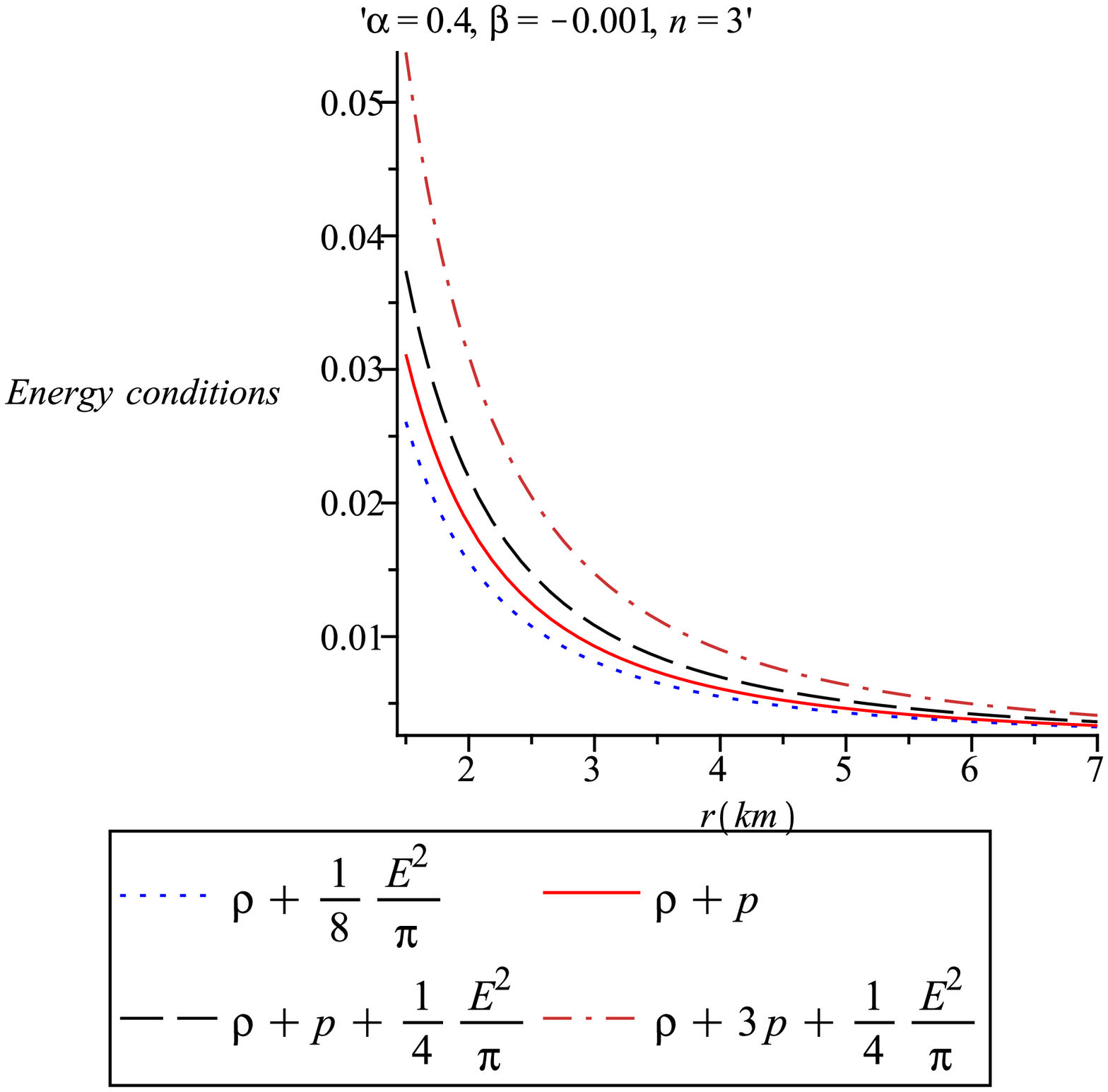}
\includegraphics[width=0.45\textwidth]{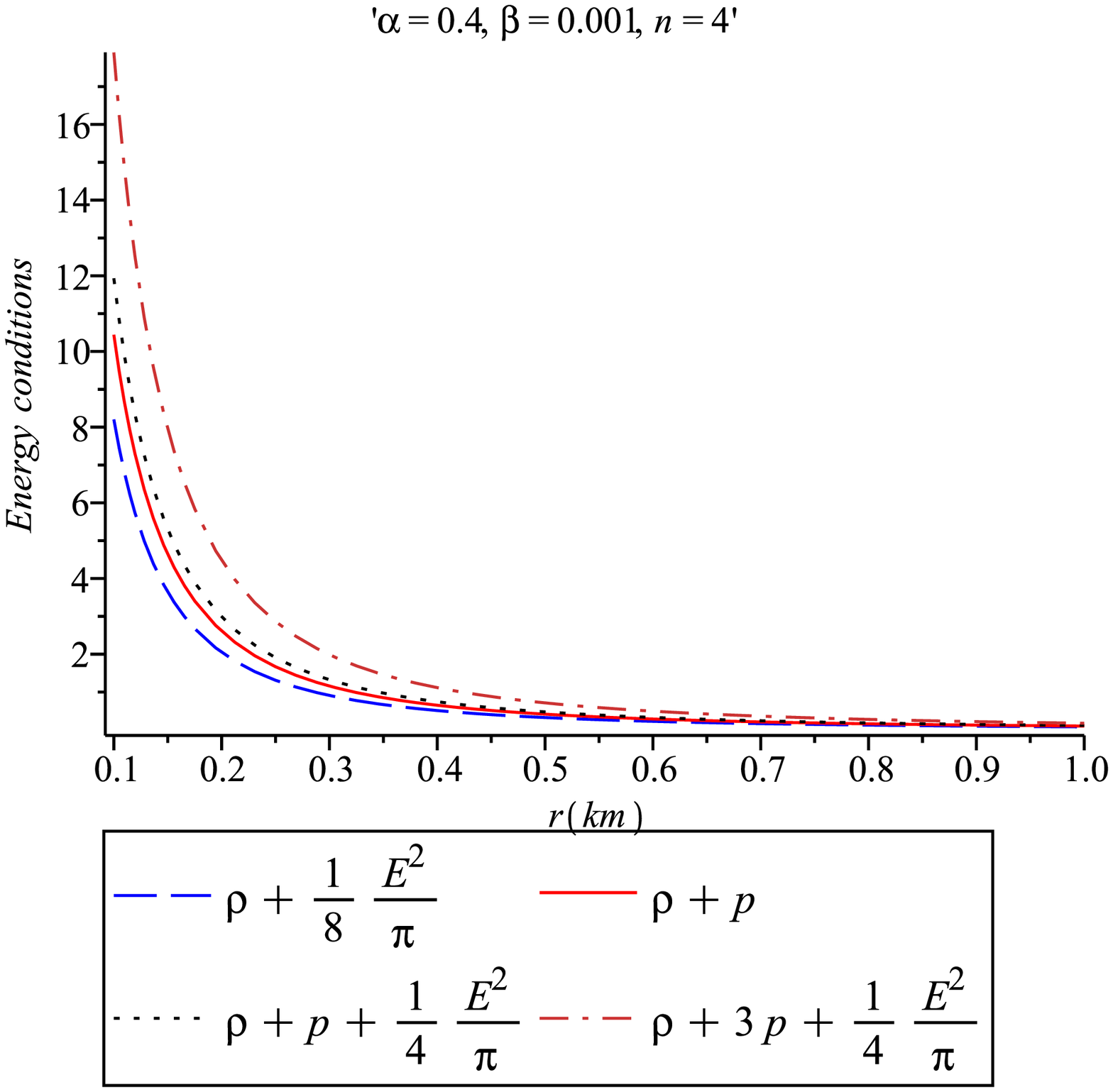}
\includegraphics[width=0.45\textwidth]{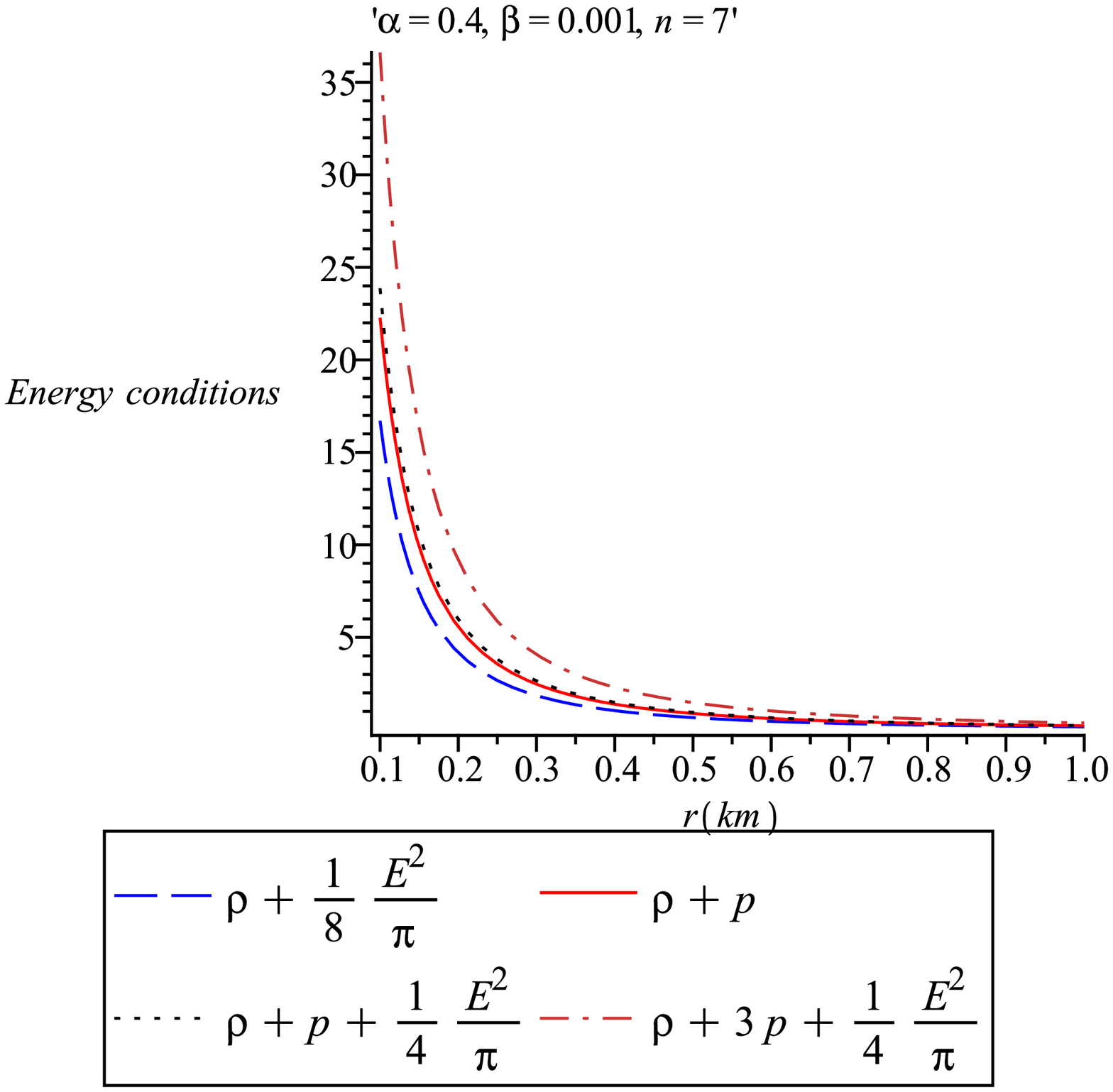}
\includegraphics[width=0.45\textwidth]{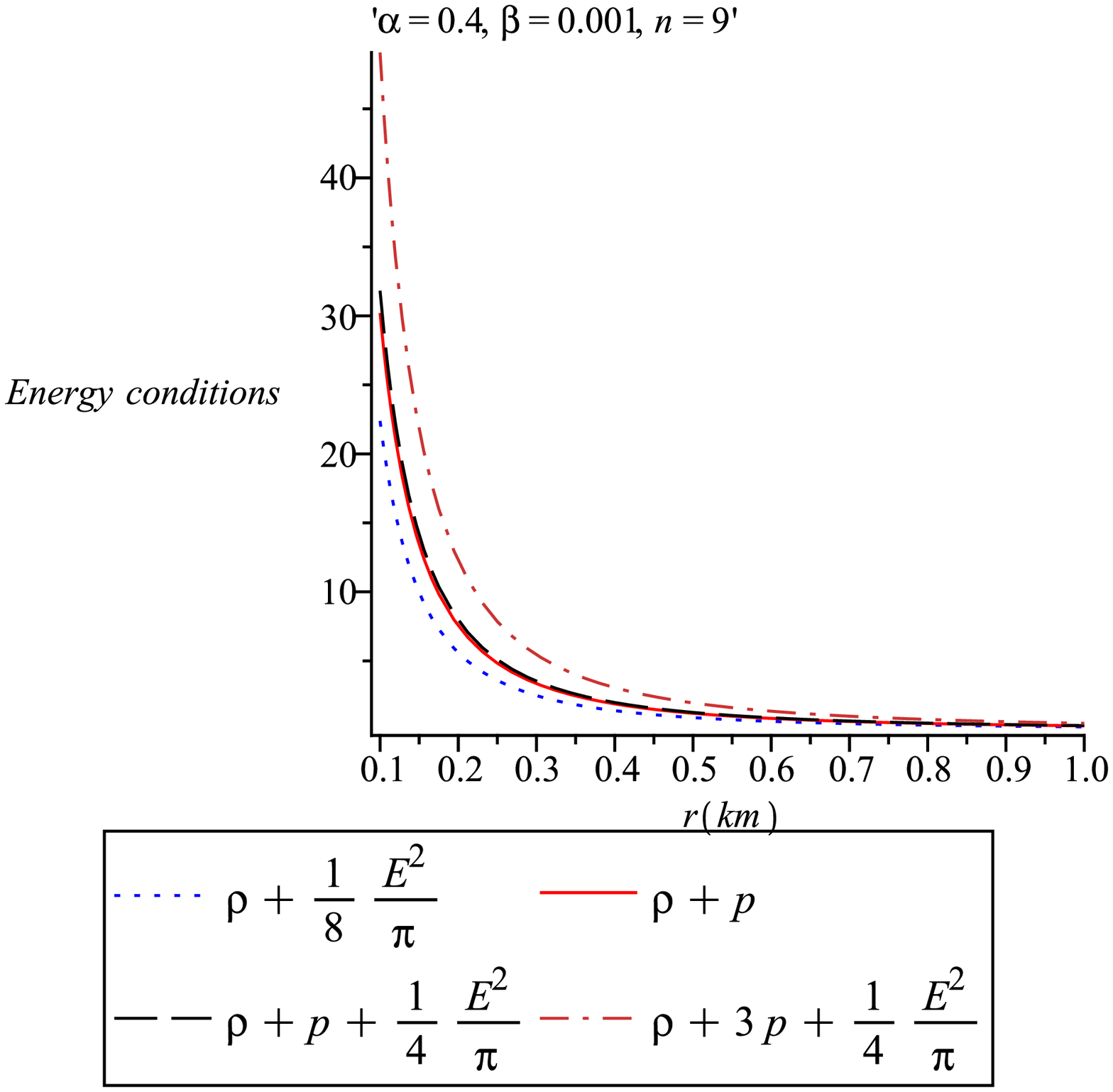}
\caption{The energy conditions are plotted against $r$ for $4D$
(upper left), $5D$ (upper right), $6D$ (middle left), $9D$ (middle
right) and $11D$ (lower)}
\end{figure}

The graphs of the Energy conditions corresponding to different
dimensions are shown in Fig. 6, which indicates that all the
Energy conditions are satisfied by proposed model of charged fluid
sphere in different dimensional spacetime.

\subsection{Compactness and Redshift}

We have already obtained the gravitational mass $m(r)$ of the
system in Eq. (30). Using this, one can easily find the
compactness factor and redshift of the star are respectively as
\begin{equation}
u(r)=\frac{m(r)}{r},
\end{equation}

\begin{equation}
z_s=(1-2u)^{-1/2}-1.
\end{equation}

The graphs of compactness factor and surface redshift corresponding to
different dimensions are given in Figs. 7 and 8.

\begin{figure}[thbp]
\centering
\includegraphics[width=0.45\textwidth]{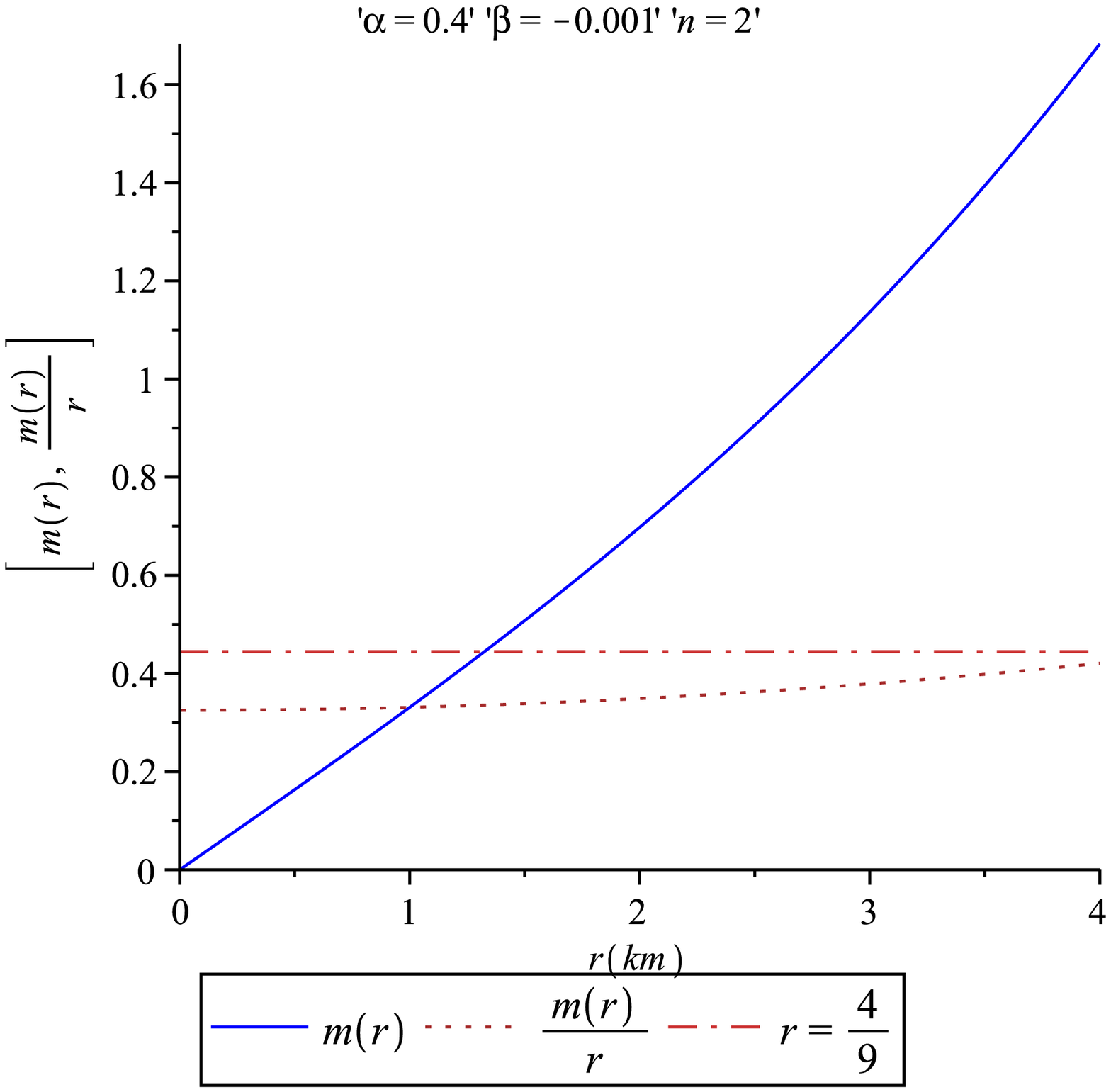}
\includegraphics[width=0.45\textwidth]{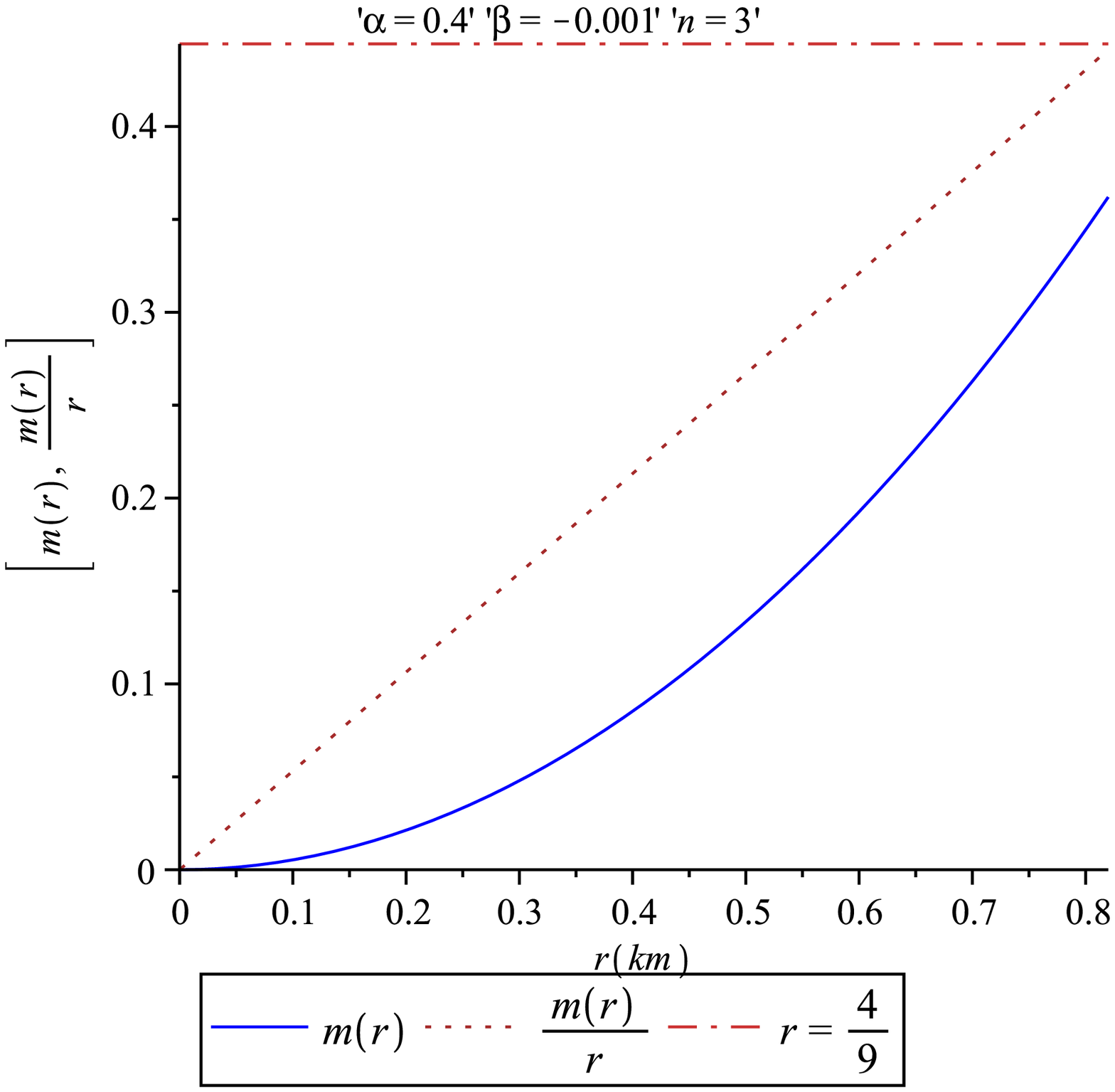}
\includegraphics[width=0.45\textwidth]{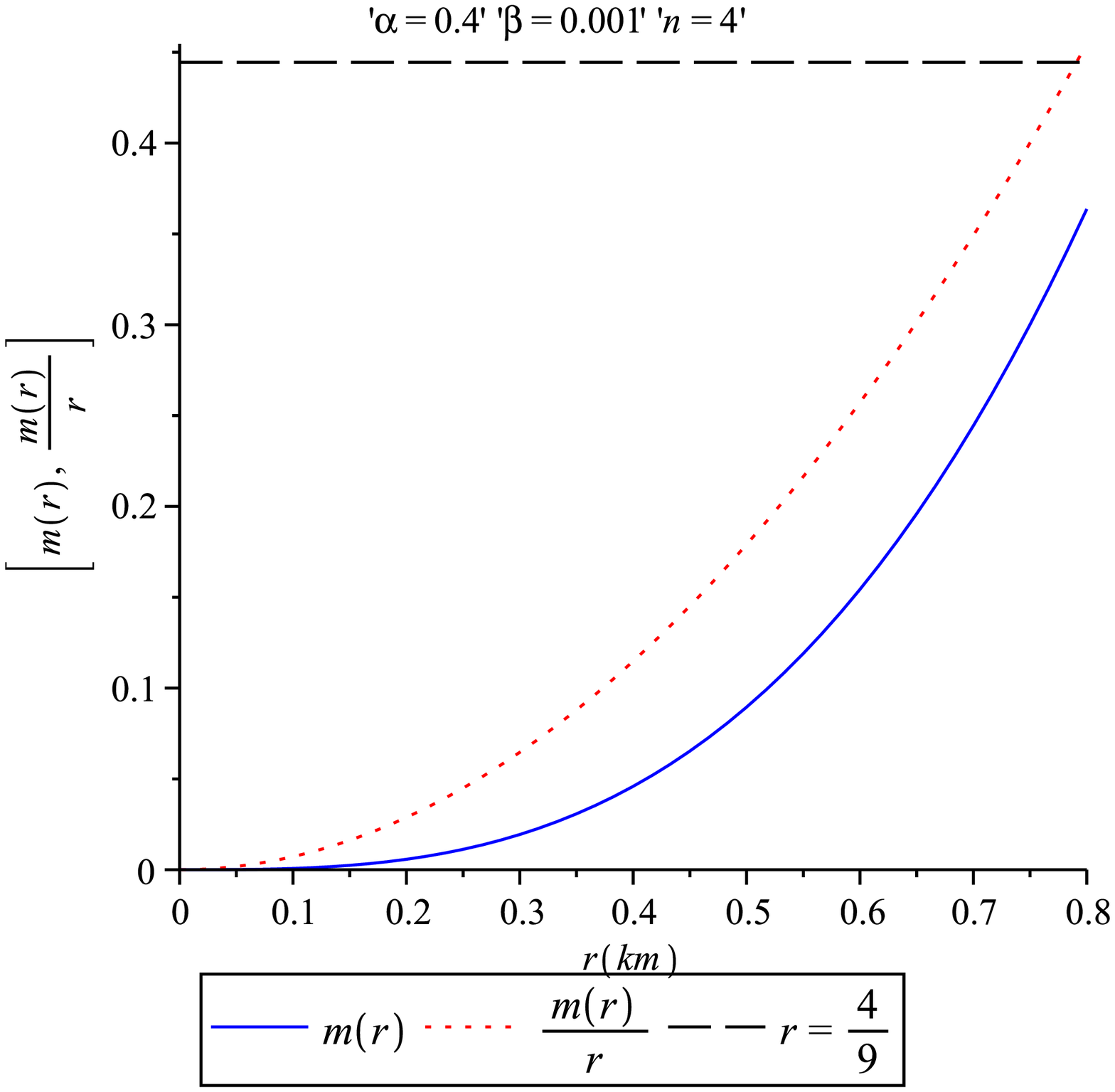}
\includegraphics[width=0.45\textwidth]{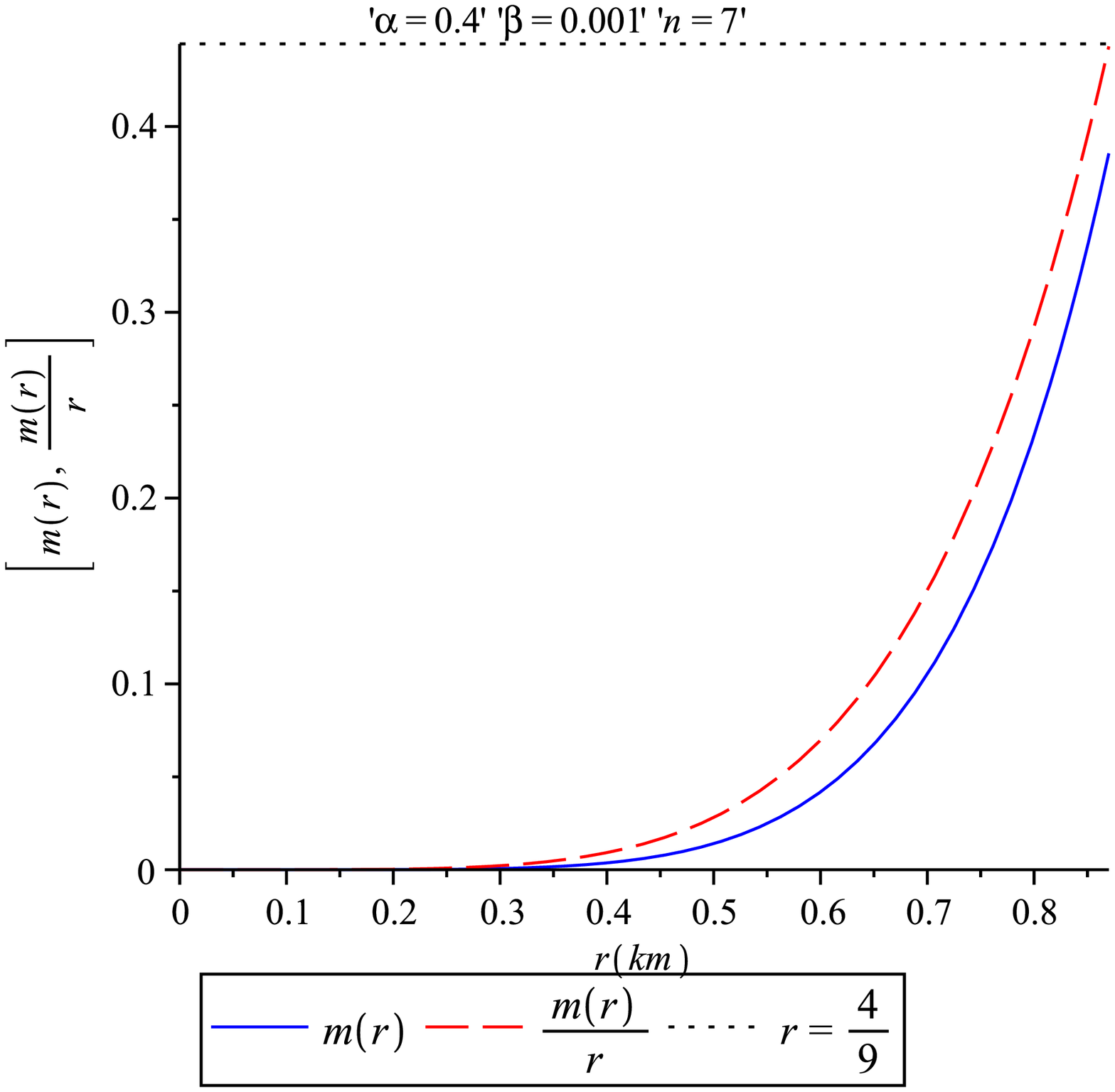}
\includegraphics[width=0.45\textwidth]{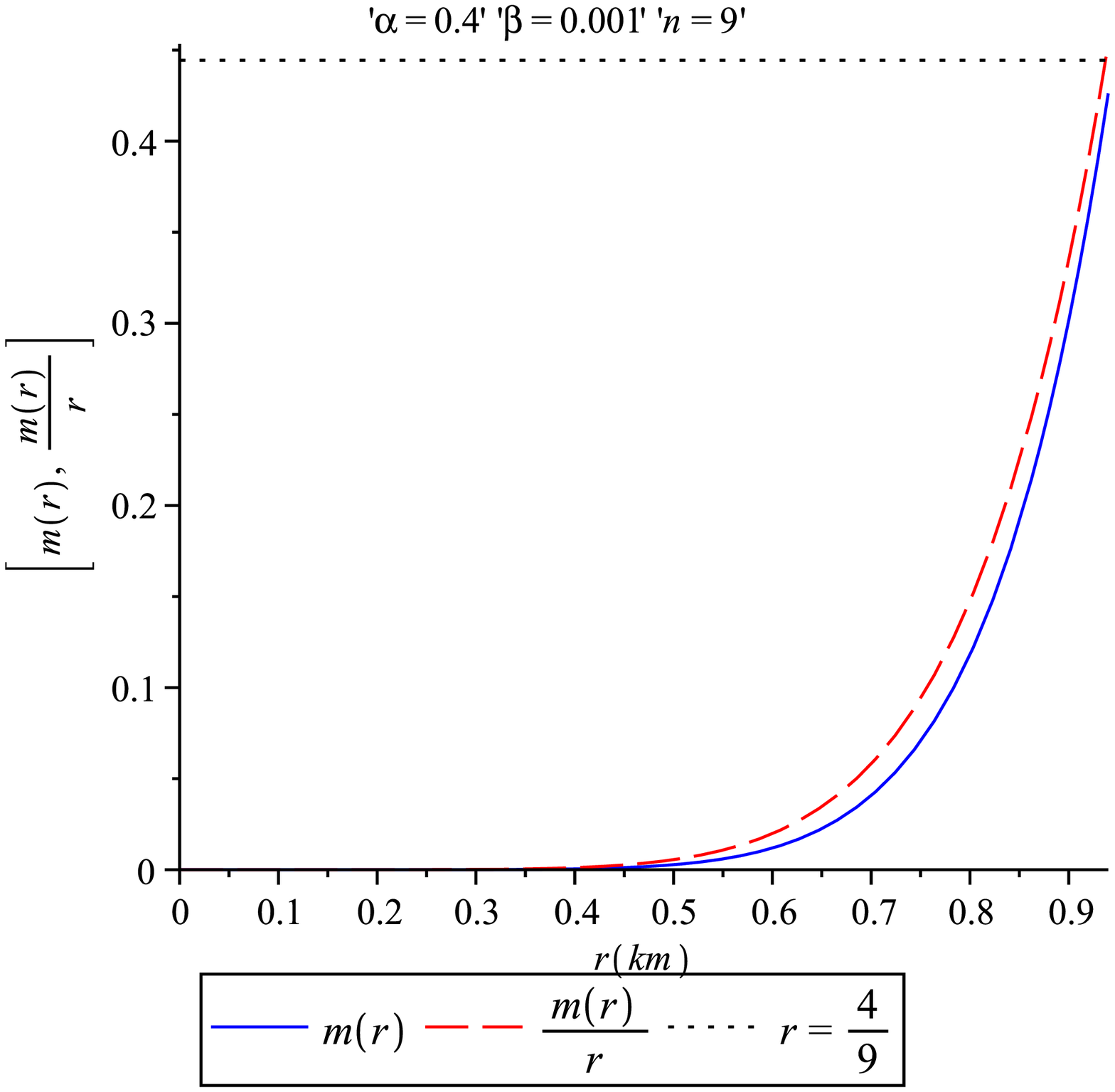}
\caption{The mass and compactness factor are plotted against $r$
for $4D$ (upper left), $5D$ (upper right), $6D$ (middle left),
$9D$ (middle right) and $11D$ (lower)}
\end{figure}

\begin{figure}[thbp]
\centering
\includegraphics[width=0.45\textwidth]{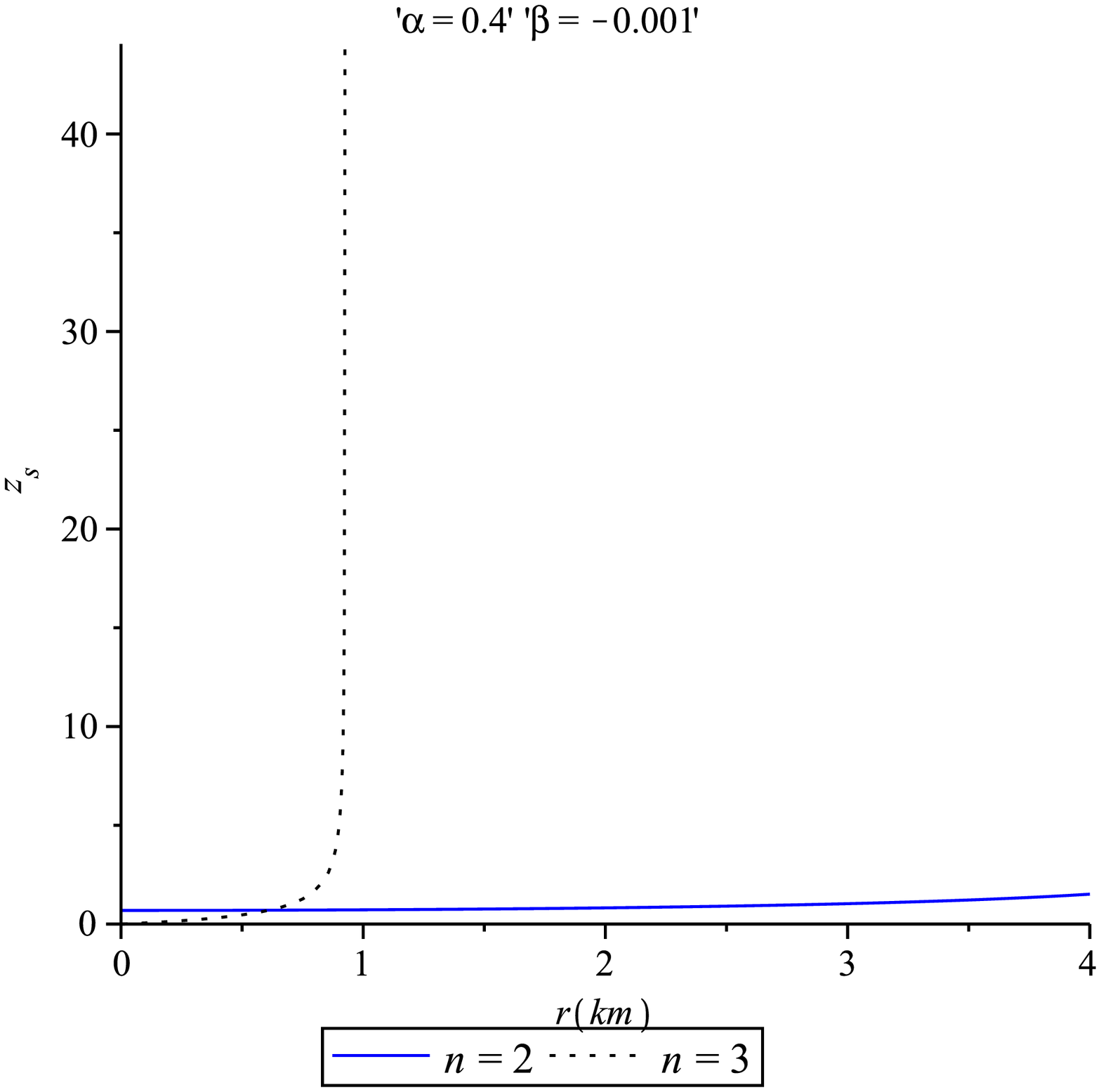}
\includegraphics[width=0.45\textwidth]{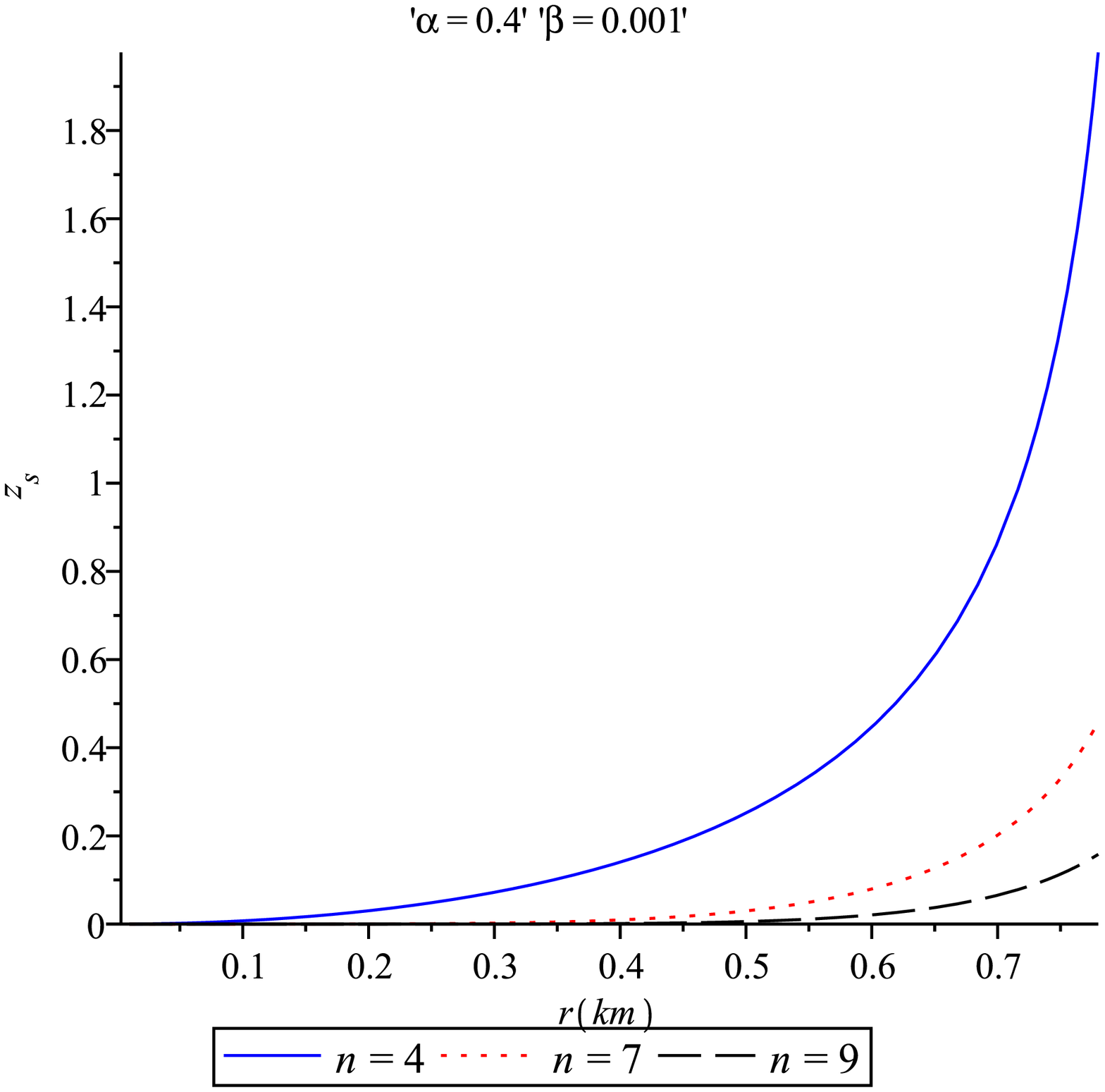}
\caption{The redshift of the star are plotted against $r$ for different
dimensions: $4D$ and $5D$ in the left panel whereas $6D$, $9D$ and $11D$ in the right panel}
\end{figure}

\section{Discussion and Conclusion}
In the present paper we have obtained a new set of interior
solutions for charged isotropic stars admitting conformal motion
in higher dimensional spacetime. Several interesting features 
have been observed which are as follows:

(1) The obtained solutions are well behaved for $r>0$. 
The matter-energy density $\rho$, fluid pressure p and
electric field intensity $E$ are all monotonically decreasing
functions of radial coordinate $r$. All the physical quantities
are mostly regular throughout the stellar configuration. 

(2) The mass of the star has been proposed (see Eq. (43)) 
in terms of the thin shell mass. 

(3) Various physical properties like the compactness factor 
and surface redshift are studied not only in the standard four 
dimensional spacetime but also in higher
dimensions. We notice that the surface redshift of the star does
not exist for the spacetime related to sixth dimensions (see Table 1).

(4) We have estimated the radii of the star in different
dimensions by means of plots (see Table 1). The radii are 
found out through the technique where the radial pressures 
meet the radial coordinate $r$. We note that for negative
value of $\beta$ radial pressures vanish at the boundary for $4D$
and $5D$ cases and for positive value of $\beta$ radial pressures
vanish at the boundary beyond five dimensions. The radii of
the star are found to be a few kilometers only for different dimensions
and masses of the stars are comparable with the mass of the sun. 
This data clearly indicates that the model represents a compact star. 
Plugging the numerical values of the physical constants $G$ and $c$ in the relevant 
expressions, we have found out mass of the star as $1.204~M_{\odot}$ and 
the surface density as $3.49\times10^{15}$~(gm/cc) in four dimensional 
background. We observe that this data as obtained in the Table~1 
is quite close to $SAX~J1808.4-3658~(SS2)$ its mass being $1.323$~$M_\odot$.

\begin{table}
\caption{Values of the physical parameters for the strange star 
candidate $SAX~J1808.4-3658~(SS2)$ related to different dimensions. 
Here the computations are shown up to the Buchdahl limit with radius $r=0.84$~km}
{\begin{tabular}{@{}ccccccc@{}} \hline Dimension & $\beta$ &
Radius $( p_r=0) $ &  Mass & Mass & $\rho_s$ &$Z_s$\\ & & (km) &
(km) & ($M_\odot$) & (gm/cc) & \\ \hline 4 & $-0.001$ & 4.14 &
1.776 & 1.204 & $ 3.49 \times 10^{15}$ & 1.637
\\ 5 &  $-0.001$ & 11 &  0.3803043117$^*$  & 0.2578334317$^*$ &
$92.61864817 \times 10^{15}$$^*$ & 2.252767160$^*$ \\ 6&  0.001&
14 &  0.4205932803$^*$ & 0.2851479866$^*$ & $139.1248886 \times
10^{15} $$^*$  & does not exist$^*$  \\ 9 &  0.001 & 7.8 &
0.3130416663$^*$ & 0.2122316382$^*$ &
$298.2815551\times10^{15}$$^*$  & 0.98160610$^*$\\ 11 &  0.001 &
7.1 & 0.1743244558$^*$ & 0.1181860717 &
$404.9245698\times10^{15}$$^*$ & 0.307505974$^*$ \\ \hline
\end{tabular}}
 \end{table}

(5) The most important result we obtain is that the model 
satisfies the Buchdahl inequality i.e. $\frac{m(R)}{R} < \frac{4}{9}$ 
for four dimensional spacetime only. For the spacetime more than four, 
Buchdahl inequality does not hold good up to the radius of the star 
in the range 4-11 km for which $p_r(r=R)=0$. However, a thourough investigation shows that 
up to the radius $0.84$~km (approx) Buchdahl inequality holds good 
for dimension more than four dimensions also. Along with this also all the
other physical characteristics are well behaved and the energy conditions
are satisfied up to radius $0.84$~km (see Table 1). 

In this regard, specifically we would like to mention that in one of our previous work~\cite{Bhar2015b} 
we obtained the radius from the plot within 1 km and, therefore, 
the above feature could not been verified. Now, it can be seen 
that if we go over 1 km then Buchdahl inequality does not 
hold good in the work of Bhar et al.~\cite{Bhar2015b} and this 
new property has been overlooked in previous all studies. 
As a summary, it reveals that though the stars under the present 
investigation have radius within the range 4-11 km, but they satisfy 
Buchdahl inequality within $0.84$~km only. 

(6) We have analyzed the TOV equations for different dimensions
which indicate that the gravitational force of the star is balanced by
the combined effect of the hydrostatic and electric forces. This 
implies that the system is in static equilibrium under these three
forces. Thus our study reveals that up to the radius for which
Buchdahl inequality holds good, the compact stars are well behaved
in higher dimensional spacetime. 

Hence on a primary stage, unlike Bhar et al.~\cite{Bhar2015b} 
and Ghosh et al.~\cite{Ghosh2015}, it seems that compact stars 
do exist even in higher dimensional spacetime as proposed 
in the current paper. However, before accepting 
this theoretical result as fact other type of investigations with 
different propositions are extmremly needed to perform.

\section*{Acknowledgments}
FR and SR wish to thank the authorities of the Inter-University
Centre for Astronomy and Astrophysics, Pune, India for providing
the Visiting Associateship.

\end{document}